\documentclass[fleqn,usenatbib]{mnras}

\usepackage{newtxtext,newtxmath}

\usepackage[T1]{fontenc}

\DeclareRobustCommand{\VAN}[3]{#2}
\let\VANthebibliography\thebibliography
\def\thebibliography{\DeclareRobustCommand{\VAN}[3]{##3}\VANthebibliography}

\usepackage{graphicx}	
\usepackage{lscape}
\usepackage{longtable}
\usepackage{bm}

\title[Galaxy rotation curve]{Galaxy rotation curve based on RGB stars from the Gaia DR3 catalogue}

\author[P. N. Fedorov et al.] {
P. N. Fedorov,$^{1}$ \thanks{E-mail: pnfedorov@gmail.com (PNF)}
A. M. Dmytrenko,$^{2}$
V. S. Akhmetov,$^{2,3}$ \thanks{E-mail: akhmetovvs@gmail.com (VSA)}
A. B. Velichko,$^{1}$
V. P. Khramtsov,$^{2}$
\newauthor
S. I. Denyshchenko,$^{1}$
I. B. Vavilova,$^{2}$
D. V. Dobrycheva,$^{2}$ 
   O. Sergijenko,$^{2}$  
A. A. Vasylenko, $^{2}$
\newauthor
O. V. Kompaniiets$^{2}$
\\
$^{1}$Institute of astronomy of V.N.Karazin Kharkiv national university, Svobody sq. 4, 61022, Kharkiv, Ukraine \\
$^{2}$Main Astronomical Observatory of the National Academy of Sciences of Ukraine, Akademik Zabolotnyi 27, Kyiv, 03143, Ukraine \\
$^{3}$INAF-Osservatorio Astrofisico di Torino, Via Osservatorio 20, Pino Torinese, Turin, I-10025, Italy
}

\date{Accepted XXX. Received YYY; in original form ZZZ}

\pubyear{2025}

\begin{document}
\label{firstpage}
\pagerange{\pageref{firstpage}--\pageref{lastpage}}
\maketitle

\begin{abstract}
In this paper, we construct a detailed circular velocity curve of the Milky Way out to $\sim\!$20 kpc based on the radial component of the Jeans equation in cylindrical coordinates, assuming an axisymmetric gravitational potential, and show its dependence on azimuth. We use only $Gaia$ DR3 data and aim to minimize the use of model data and various assumptions.

To build the rotation curves, we used a sample of 4,547,980 RGB stars with measured spatial velocities, covering the Galactic plane in the range of Galactocentric cylindrical coordinates $150^\circ<\Theta<210^\circ$ and $0<R<20$ kpc. We exclude systematics in the data that may arise from neglecting higher-order moments of the velocity distribution and their dispersions, as well as due to random measurement errors in $Gaia$.

At the Sun’s distance, the circular velocity $V_{\rm c}(R_0)$ turned out to be ($229.63\pm0.30$) km s$^{-1}$, which is in 
good agreement with many previous estimates. The average slope of the circular velocity is $(\sim-2.29\pm0.05)$ km 
s$^{-1}$ kpc$^{-1}$ obtained in the range of $R$ from 6 to 20 kpc and $\Theta$ from 150 to 210 degrees. The determined circular velocity curve has some peculiarities in behavior near $R\sim$13 and 18 kpc,
but in general it does not contradict the results of other authors up to distances where our statistics are reliable.

\end{abstract}

\begin{keywords}
methods: data analysis--proper motions--stars: kinematics and dynamics--Galaxy: kinematics and dynamics--solar neighborhood.
\end{keywords}

\section{Introduction}

The circular velocity curve $V_{\mathrm{c}}(R)$ of the Milky Way is usually
constructed along the Galactocentric direction $R$ passing through the Galactic
centre and the Sun, using the axisymmetric Jeans equation that links the
gravitational potential to the stellar kinematics encoded in the
velocity--dispersion tensor. This procedure implicitly assumes that
$V_{\mathrm{c}}(R)$ represents a global characteristic that may be extrapolated
to the Galaxy as a whole. The justification for extrapolating the rotation
curve throughout the Milky Way is the assumption of axial symmetry of the
gravitational potential entering the Jeans equation. In practice, however, the
available kinematic information covers only about one third of the Galactic
disc, as observations cannot be obtained across the entire Galaxy. If the
rotation curve varies with azimuth $\Theta$, then the classical approach to
estimating the Galactic mass using an azimuthally averaged $V_{\mathrm{c}}(R)$
loses strict physical meaning.

The key assumptions underlying the Jeans equation, namely axial symmetry and
dynamical equilibrium of the stellar population under study, are known to be
violated, as demonstrated in numerous works
\citep[][]{Fedorov2023,Akhmetov2024,Dmytrenko2023,Denyshchenko2024,Dmytrenko2025}.
These studies show that the distribution of azimuthal velocity components of
stellar centroids in the Galactic plane exhibits significant variations at
different azimuthal angles, and that the Milky Way is neither strictly
axisymmetric nor in dynamical equilibrium \citep[][]{Vasiliev2021, Antoja2018}. Consequently, the determination of
the circular velocity curve via the Jeans equation, and thus the inference
of the Galactic mass, becomes strongly model-dependent. Current estimates of the
Milky Way mass range from $\sim4\times10^{11}~{\rm M_\odot}$ to
$1.5\times10^{12}~{\rm M_\odot}$, i.e. differing by more than a factor of three.
This discrepancy arises not from an experimental error but from incomplete data and
model assumptions.

A robust determination of $V_{\mathrm{c}}(R)$ requires precise kinematics for a
sample containing a large number of stars with well-measured space velocities,
ensuring statistically reliable results. Gaia data releases, being the largest
and most precise available, have often been used to determine the circular
velocity not only within the Solar neighbourhood but also at distances up to
20--30~kpc
\citep{ Mroz2019,Lopez-Corredoira2019,Eilers2019,Chrobakova2020,Wang2023}.
However, two serious obstacles complicate this task. First, the precision of
parallax-based distances deteriorates rapidly, especially beyond heliocentric
distances of $\sim5$~kpc. Second, reaching such distances requires the use of
intrinsically luminous stars, as only such stars are observable by
\textit{Gaia} at large ranges. Red giant branch (RGB) stars are therefore widely
used as tracers in rotation-curve studies. Although RGB stars are useful
tracers, their observed velocities are affected by asymmetric drift, which
systematically shifts the inferred circular velocity and is nontrivial to
correct for.

Significant efforts have been devoted to mitigating the first limitation by
improving the accuracy of \textit{Gaia} parallaxes. For example,
\citet{Lindegren2021} proposed parallax-correction functions.
\citet{Eilers2019} used spectrophotometric parallaxes derived from APOGEE DR14
spectroscopy and multi-band photometry, while \citet{Mroz2019} utilised the
Cepheid catalogue of \citet{Skowron2019}. \citet{Bailer-Jones2021} introduced the
so-called photogeometric distances combining a Galactic prior with astrometric
data, and \citet{Lopez-Corredoira2019} proposed a distance-refinement method
based on statistical deconvolution of parallax errors using the Lucy inversion
algorithm \citep{Lucy1974}.

To reduce the impact of asymmetric drift, many works apply restrictions such as
excluding stars with $|Z|>1$--2~kpc, adopting $\alpha$-element enhancement cuts to
separate thin- and thick-disc populations, or removing halo stars. These
criteria reduce the influence of asymmetric drift but do not ensure its complete
elimination. Moreover, such restrictions reduce the number of available stars,
particularly at large distances, lowering the statistical robustness of the
results.

The latest \textit{Gaia} release, DR3 \citep[][]{Antoja2021,Drimmel2023}, contains
32.2 million stars with radial velocities brighter than $G_{\rm RVS}\!\sim\!14$,
with median radial-velocity precision ranging from 1.3~km\,s$^{-1}$ at
$G_{\rm RVS}\!\sim\!12$ to 6.4~km\,s$^{-1}$ at $G_{\rm RVS}\!\sim\!14$.
Random proper-motion uncertainties are 0.02--0.03~mas\,yr$^{-1}$ for $G<15$,
0.07~mas\,yr$^{-1}$ at $G=17$, 0.5~mas\,yr$^{-1}$ at $G=20$ and
1.4~mas\,yr$^{-1}$ at $G=21$. Parallax uncertainties have been reduced by a
factor of 1.25. Using these data and the above mentioned distance-refinement
approaches, circular-velocity curves have been extended to
$R\sim25$--30~kpc
\citep[][]{Ou2024,Zhou2023,Wang2023,Jiao2023,Koop2024}. However, no consensus has
yet been reached regarding the behaviour of the rotation curve at large radii:
some works report a sharp, nearly Keplerian decline beyond $\sim$15--25~kpc,
while others find an almost flat or slowly declining profile out to $\sim$30~kpc
\citep[e.g.][]{Chrobakova2020}. The reasons for discrepancies at large distances 
may be the non-stationarity of stellar systems, inaccurate selection of tracers, 
asymmetries, and inaccurate distances \citep[][]{Wang2023,McMillan2022}. 
The applicability of the Jeans equation, with its
assumptions of stationarity and symmetry, has therefore come under close examination
\citep{Koop2024}, and many works introduce corrections or explicitly
model disequilibria.

Despite these challenges, refining the Milky Way circular-velocity curve
$V_{\mathrm{c}}(R)$ remains crucial, as it provides a practically useful
approximation and plays an important role in many astrophysical applications.
For instance, the shape of $V_{\mathrm{c}}(R)$ is used when comparing the Milky
Way with external disc galaxies, since RC shapes reflect galaxy type. Most
spiral galaxies display flat rotation curves at large radii, although some show
declines, especially when traced with different indicators or at very low
surface densities \citep{Puglisi2023}. Such comparisons help assess whether a
possible downturn in the Milky Way rotation curve is unusual or typical.

The goal of the present study is to determine the Galactic rotation curve along
different azimuthal directions using \textit{Gaia}~DR3 data alone, without any
model-dependent inputs. A second objective is to derive the Galactic circular
velocity using the Jeans equation with a minimal set of assumptions, employing
all relevant velocity and dispersion moments corrected for \textit{Gaia}
measurement errors. The analysis is performed in spherical regions (bins) of
radius 1~kpc, placed every 100~pc, enabling the construction of maps of all
quantities needed to evaluate the Jeans equation after removing potential
systematics from the stellar velocity field.

\section{Data and Sample}

In this study, we use RGB stars. Following
\citet{Vallenari2023}, we identify RGB stars using the stellar-parameter
criteria $\log g < 3.0$ and $3000~{\rm K} < T_{\rm eff} < 5200~{\rm K}$,
as provided in \citet{Andrae2023}. This selection yields a total of
10\,976\,663 sources. Among them, 5\,913\,442 stars have radial velocities
reported in \textit{Gaia}~DR3.

As in our previous works \citep{Fedorov2023,Dmytrenko2025}, we remove from
the sample all stars satisfying any of the following criteria:
\begin{align}
&{\tt RUWE} > 1.4, \nonumber \\
&\frac{\varpi}{\sigma_{\varpi}} < 5,\quad \text{if } R \leq 15~{\rm kpc}, \nonumber \\
&\varpi > 0,~~~~\quad \text{if } R > 15~{\rm kpc}, \nonumber \\
&|V_Z| > 100~{\rm km\,s^{-1}}.
\label{eq:sel_criteria}
\end{align}

The parallaxes of stars in the sample were corrected following the
prescriptions of \citet{Lindegren2021}, using the Z5 and Z6 parallax-bias
terms. For stars with absolute magnitudes $M_G$ between 9 and 13, we also
correct the proper motions using the recommendations of
\citet{Cantat-Gaudin2021}. In addition, we remove the magnitude-dependent
trend in radial velocities according to the prescription of D.~Katz
(\textit{Gaia} DR3 documentation). For stars with {\tt rv\_template\_teff}
$< 8500~{\rm K}$, we subtract from the reported \textit{Gaia} radial velocity
the correction
\begin{equation}
V_{\rm rcorr} =
\begin{cases}
0, & \text{if } {\tt grvs\_mag} < 11, \\
0.02755\,g^2 - 0.55863\,g + 2.81129, & \text{if } g \ge 11,
\end{cases}
\end{equation}
where $g \equiv {\tt grvs\_mag}$ and $V_{\rm rcorr}$ is given in
km\,s$^{-1}$.

After applying the quality cuts and corrections mentioned above, the final RGB
sample contains 4\,547\,980 stars with full space velocities. This sample
is used in all subsequent calculations, including the determination of both
the Galactic rotation curve and the circular velocity curve.

\section{Coordinate Systems}

We introduce a local, right-handed Cartesian coordinate system
$(x,y,z)$ at any arbitrary point in the Galactic disc, analogous to the
Galactic rectangular system $(X,Y,Z)$. This is possible because for
any given point (e.g., a star) and for the stars located in its neighbourhood the spatial
coordinates and velocity components are known. The transformation from the
Galactic Cartesian system to a local Cartesian system is equivalent to
placing a fictitious observer at the position corresponding to the Galactic
coordinates of the local system's origin.

\begin{figure}
    \centering
    \includegraphics[width=0.95\linewidth]{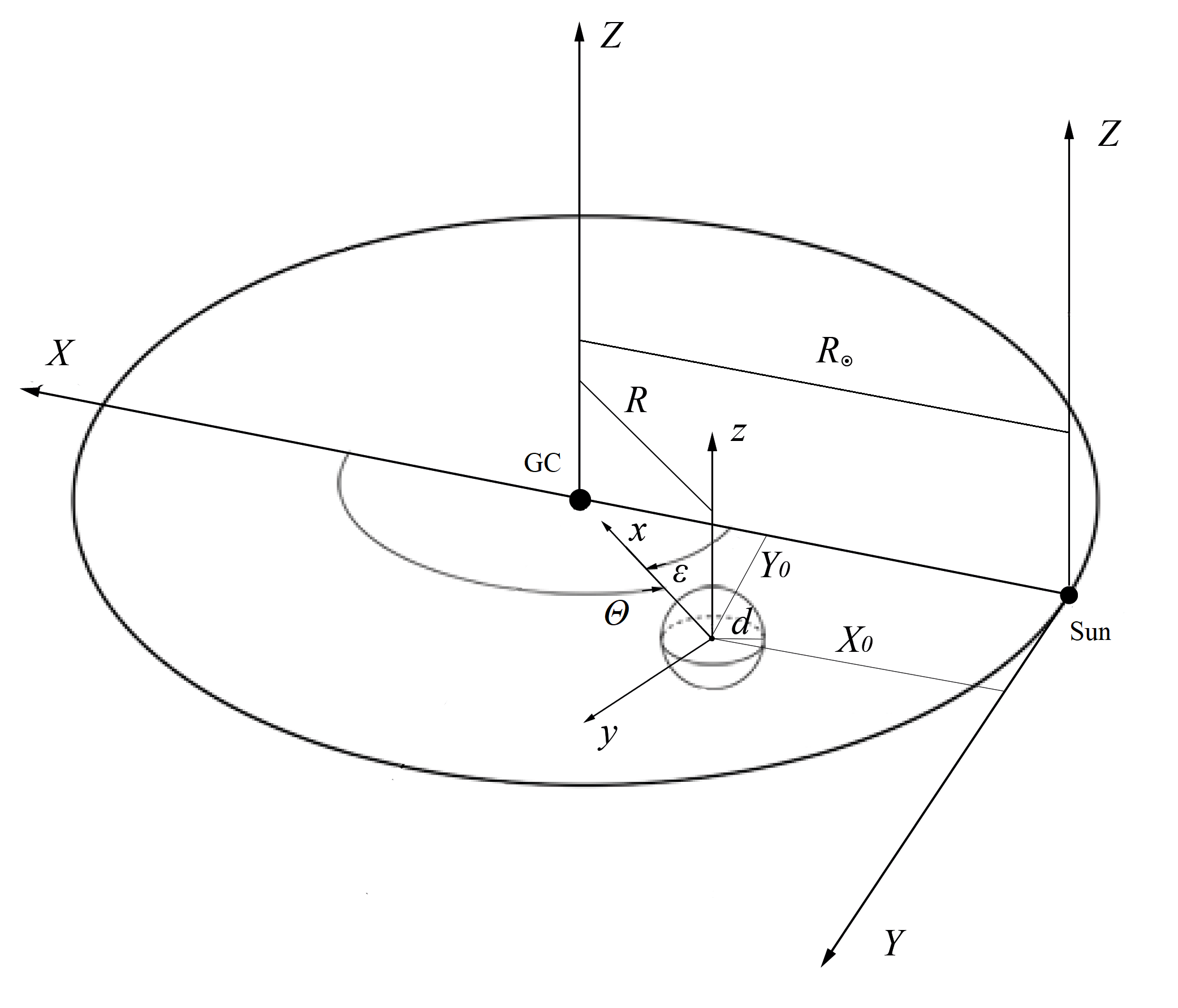}
    \caption{The link between the Galactocentric, Galactic and local coordinate systems.}
    \label{fig:MW_scheme}
\end{figure}

To ensure consistency between the $(X,Y,Z)$ Cartesian system and the
Galactocentric cylindrical system $(R,\Theta,Z)$, Fig. \ref{fig:MW_scheme}, we adopt the Solar distance
from the Galactic centre of $R_\odot = 8.122~{\rm kpc}$ obtained by
\citet{Abuter2021}, and the Solar velocity components
$(U_\odot, V_\odot, W_\odot) = (11.1,\; 245.8,\; 7.8)~{\rm km\,s^{-1}}$
\citep{Reid2004}.

The local Cartesian coordinate system is defined in such a way that
its $x$-axis always points from the centroid towards the Galactic rotation
centre, the $y$-axis coinsides with the direction of Galactic rotation,
and the $z$-axis is parallel to the direction from the Galactic center to its north pole.

The `observation points' from which the fictitious observer samples the
stellar velocities are defined as follows. For stars with relative parallax
errors smaller than 20 per cent and Galactocentric distances less than
15~kpc, we construct spherical volumes of radius 1~kpc. The centres of these
spheres are placed at the nodes of a rectangular grid located in the Galactic
mid-plane, i.e. all grid nodes satisfy $Z=0$. The grid spacing in both $X$ and
$Y$ directions is 100~pc.
For stars at Galactocentric distances larger than 15~kpc, we apply the
condition $\varpi>0$. All stars
within each spherical region are used to form a stellar system whose
centroid velocity is defined as the mean velocity of the stars in that
sphere. Thus, our sample was always limited by the condition $|Z| < 1$ kpc.
The same stellar systems are constructed in the planes $Z=-1$~kpc,
$Z=-0.5$~kpc, $Z=0.5$~kpc and $Z=1$~kpc. This allows us to determine the
velocities above and below the Galactic plane, specifically at
$Z=\pm0.5$~kpc.

\section{Velocity Maps and Determination of the Galactic Rotation Curve}

The rotation curves $V_{\Theta}(R)$ derived from
observations of externally perturbed stellar systems do not coincide with the
circular velocity $V_{\rm c}(R)$ because of the presence of asymmetric drift.
Recovering the true circular velocity requires applying an appropriate
correction, which is often difficult to determine reliably. Even when such a
correction can be attempted, it is not derived from direct measurements, but
rather from model reconstructions based on azimuthally averaged line-of-sight
velocities and assumptions about the shape of the dispersion profile and the
degree of anisotropy. Therefore, in a number of applications, for example
in the task of identifying Milky Way analogues \citep[e.g.][]{Vavilova2024, Kompaniiets2025},
it is more appropriate to compare the directly observed rotation curve
$V_{\Theta}(R)$ of the Milky Way disc with the always-observable rotation
curves of external edge-on galaxies.

Because our goal is to study the velocity field over an extended region of the
Galactic plane, accurate distance estimates are required. In practice, the
distances (parallaxes) in \textit{Gaia} DR3, particularly for stars with large
parallax uncertainties, must be refined. To minimize reliance on model
assumptions, we use approaches based as far as possible on purely observational
or geometric reasoning. We examine several methods for obtaining the most
accurate distance estimates, including direct inversion of \textit{Gaia}
parallaxes and the Lucy deconvolution method \citep{Lucy1974}. Direct
inversion produces significantly distorted Galactocentric distances already at
$R \gtrsim 12$--13~kpc.

The Lucy inversion technique has been applied in several studies,
including \citet{Lopez-Corredoira2019,Chrobakova2020,Wang2023}. According to
the authors, the method reduces parallax errors, increases the number of
sources used, and extends the distances that can be reached. In other works,
spectrophotometric distances were employed: \citet{Eilers2019} and
\citet{Ou2024} used such distances for more than 30\,000 RGB stars.
\citet{Zhou2023} used a similar approach for $\sim$54\,000 thin-disc stars,
although with a different set of priors. Nearly all other approaches rely, to
some extent, on priors and approximate Galactic models to infer distance
probabilities.

Our analysis showed that, for our purposes, the most suitable distance
estimates are those obtained using the parallax-bias correction proposed by
\citet{Lindegren2021}, i.e.
\begin{equation}
\varpi_{\rm corr} = \varpi_{\rm obs} - Z_5 - Z_6,
\end{equation}
where $Z_5$ and $Z_6$ are the recommended color- and magnitude-dependent bias
terms.

To determine the post-correction parallax uncertainty, we adopt
\begin{equation}
\sigma^2_{\varpi,{\rm corr}} = \sigma^2_{\varpi,{\rm obs}} +
\sigma^2_{\rm sys},
\end{equation}
where $\sigma_{\varpi,{\rm obs}}$ is the formal \textit{Gaia} DR3 parallax
uncertainty, and $\sigma_{\rm sys} \approx 10~\mu{\rm as}$ represents the
external systematic uncertainty
\citep{MaizApellaniz2021,MaizApellaniz2022,Ding2025}.

The heliocentric distance is computed as
\begin{equation}
d = \frac{1}{\varpi_{\rm corr}},
\end{equation}
and its uncertainty follows from error propagation:
\begin{equation}
\sigma_d = \frac{\sigma_{\varpi,{\rm corr}}}{\varpi_{\rm corr}^2}.
\end{equation}

Using these values, and transforming from $(d,\mu_{\alpha},\mu_{\delta},V_r)$
to the Galactocentric cylindrical components $(V_R,V_{\Theta},V_Z)$, as well
as converting $(\sigma_d,\sigma_{\mu_\alpha},\sigma_{\mu_\delta},\sigma_{V_r})$
into $(\sigma_{V_R},\sigma_{V_{\Theta}},\sigma_{V_Z})$, we obtain individual
Galactocentric velocities and associated uncertainties for each star.

\begin{figure}
    \centering
    \includegraphics[width=1\linewidth]{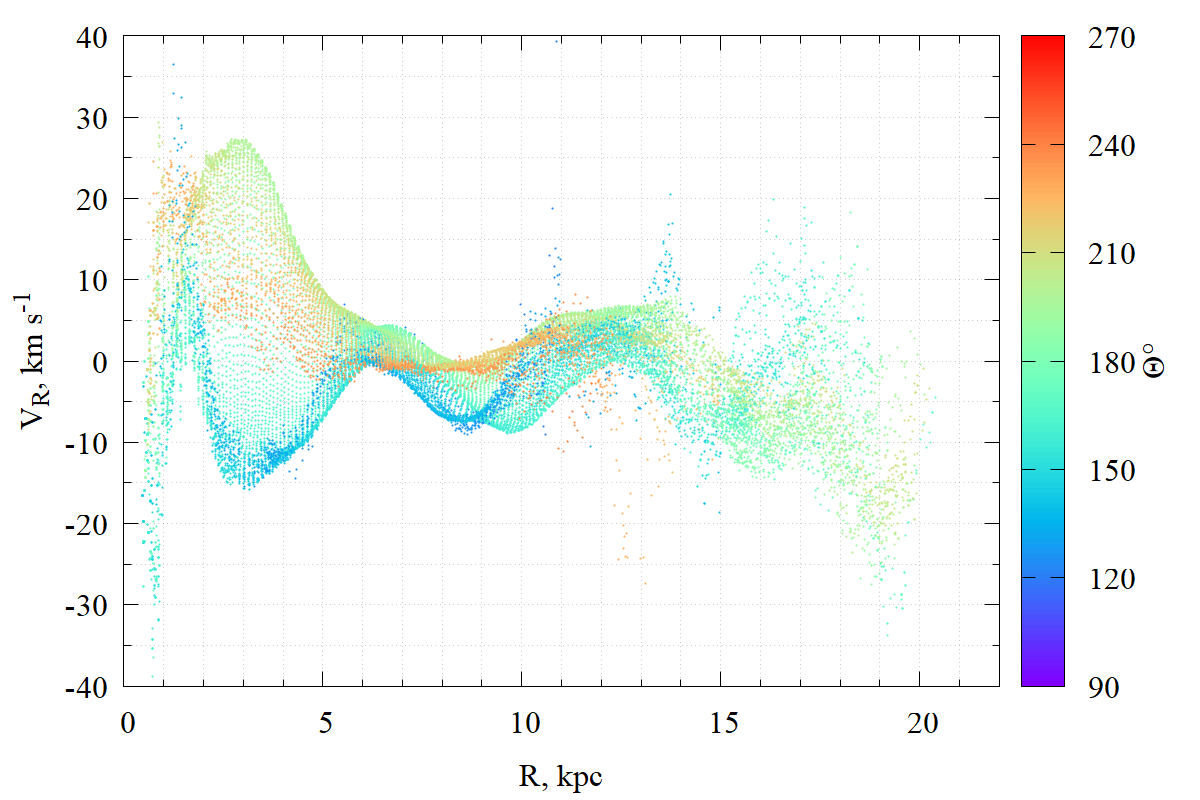}
    \caption{The centroid velocity component $V_R$ as a function of Galactocentric coordinates.}
    \label{fig:VR}
\end{figure}

\begin{figure}
    \centering
    \includegraphics[width=1\linewidth]{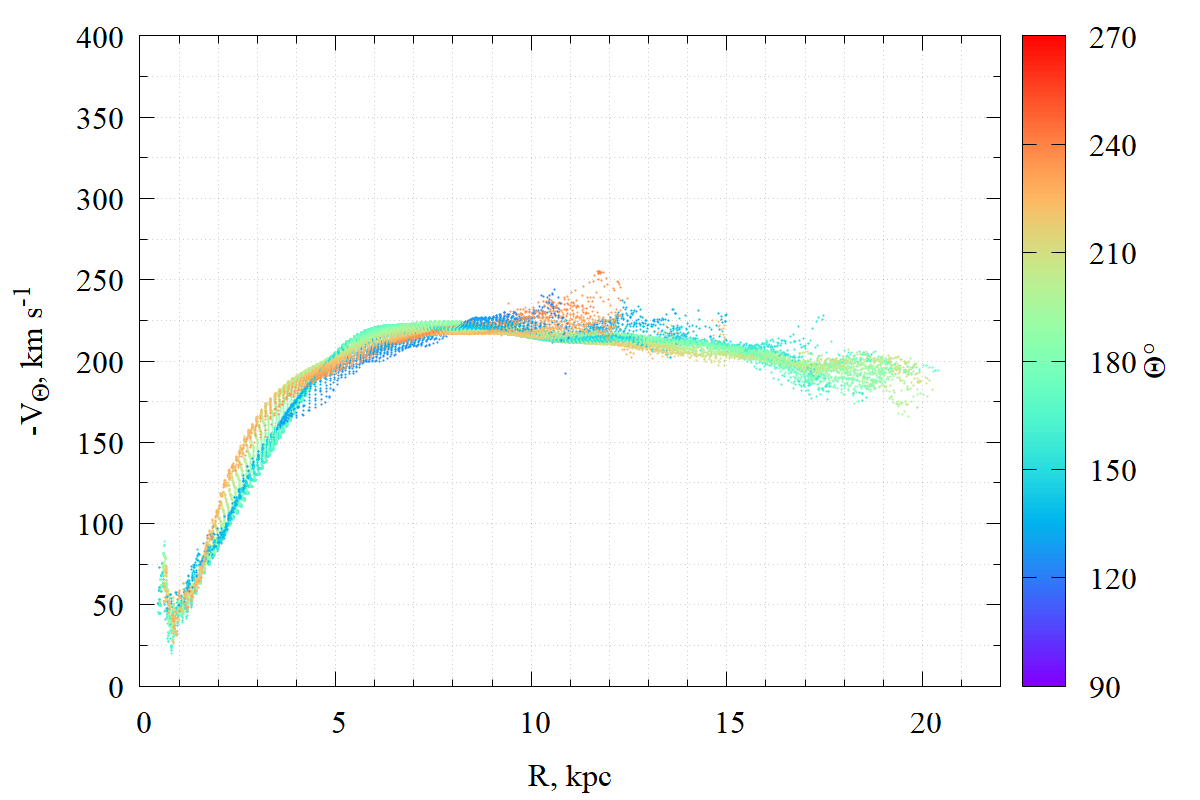}
    \caption{The centroid velocity component $V_\Theta$ as a function of Galactocentric coordinates.}
    \label{fig:VTheta}
\end{figure}

\begin{figure}
    \centering
    \includegraphics[width=1\linewidth]{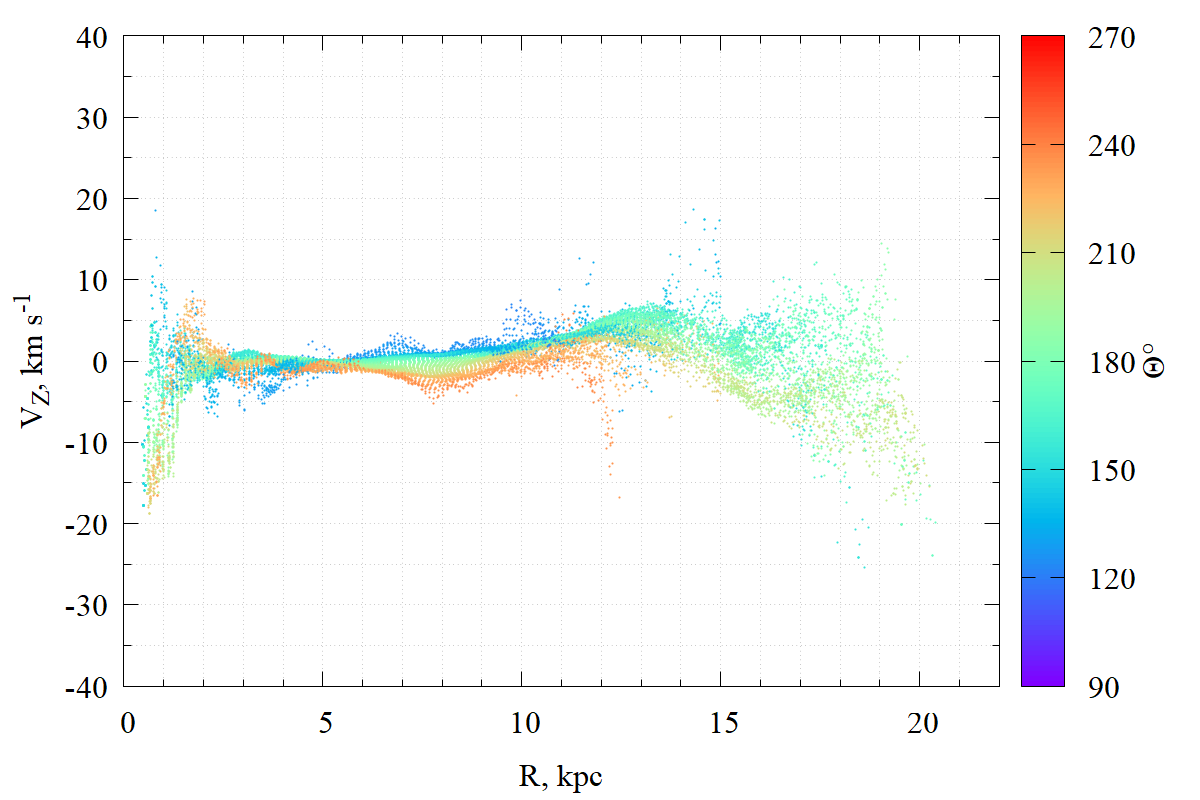}
    \caption{The centroid velocity component $V_Z$ as a function of Galactocentric coordinates.}
    \label{fig:VZ}
\end{figure}

As in \citet{Fedorov2023}, the centroid velocities $V_R$, $V_{\Theta}$, and
$V_Z$ and the nine remaining kinematic parameters of the linear
Ogorodnikov--Milne model \citep{Ogorodnikov1965} were computed from the
individual stellar velocities contained within each bin (a sphere of radius
1~kpc). The resulting Galactocentric cylindrical velocity components of the
centroids are shown in Figs. \ref{fig:VR}, \ref{fig:VTheta}, \ref{fig:VZ} as functions of Galactocentric
coordinates, after applying the parallax corrections described above.



\section{Determination of the Galactic Circular Velocity from the Jeans Equation}

In this work, we derive the Galactic rotation curve using purely kinematic
quantities that are free from model assumptions. Nevertheless, because the
Milky Way is observed from the inside, the reconstruction of the circular velocity
$V_{\rm c}(R)$ inevitably requires modeling. From the interior of the Galaxy
one can determine the full three-dimensional velocities of individual stars,
making it possible to recover, via the Jeans equation, a rotation curve of
unmatched precision and spatial detail out to $\sim$25~kpc
\citep{Ou2024,Zhou2023,Wang2023,Jiao2023,Koop2024}. Such rotation curves
provide not only constraints on the total mass of the Galaxy, but also serve as
key diagnostics when identifying Milky Way analogues. However, as recently
shown by \citet{Koop2024}, axisymmetric stationary Jeans modeling becomes
unreliable at large radii, requiring caution when interpreting the circular
velocity in the outer regions.

To derive the circular velocity curve and study its azimuthal variations, we
use the same stellar sample and the same selection criteria as described in the
previous section, adding a few additional requirements. In a given gravitational
potential $\Phi(R,z)$ one can compute the velocity an object would have if it were
moving on a perfect circular orbit of radius $R$, where the gravitational force
is balanced by the centrifugal force. To determine this velocity, we use the
radial Jeans equation in cylindrical coordinates $(R,\Theta,z)$
\citep{Binney2008}:

\begin{equation}
\label{eq:radialjeans}
\frac{\partial (\nu\langle \varv_R^2\rangle)}{\partial R}
+
\frac{\partial (\nu\langle \varv_R \varv_Z\rangle)}{\partial z}
+
\nu\left(
\frac{\langle \varv_R^2\rangle - \langle \varv_\Theta^2\rangle}{R}
\right)
=
-\nu\,\frac{\partial \Phi}{\partial R},
\end{equation}
where $\nu(R,Z)$ is the tracer density, and angle brackets denote ensemble
averages.

The radial gravitational force in equation~(\ref{eq:radialjeans}) is related to
the circular velocity by
\begin{equation}
\label{eq:vcgeneral}
V_{\rm c}^2(R,z) = R\,\frac{\partial \Phi}{\partial R}.
\end{equation}
This quantity is the mathematically defined speed of an ideal circular orbit at
a given radius. It is not a directly observable velocity, but a model-based
quantity tied to the mass distribution of the Galaxy (including dark matter).
The term `ideal' emphasizes that this is not a measurable quantity, but the velocity which the star would have if it were moving in a circular orbit without any perturbations. In reality, the observed velocity and the circular velocity do not coincide. The difference between them contains information about the Galactic dynamics and the mass distribution. Equation \ref{eq:radialjeans} shows that the radial force is balanced by the presence of the second moments of the velocity distribution and the tracer density.

Equation \ref{eq:radialjeans} assumes that $\Phi(R, z)$ is the galactic axisymmetric potential, $R$ is the radius of the Galaxy, $\varv_R$ is the radial velocity, $\varv_\Theta$ is the azimuthal velocity, $\varv_Z$ is the vertical velocity, and angle brackets $\langle\cdot\rangle$ denote the average value of a given quantity.


Following previous works of \citet{Ou2024,Bland-Hawthorn2016}, we represented the population density distribution of tracers $\nu(R, z) $ by an analytical function of the form:
\begin{equation}
\label{eq:densitynu}
\nu(R,z) \propto
\exp\!\left(-\frac{R}{h_R}\right)
\exp\!\left(-\frac{|z|}{h_z}\right),
\end{equation}
with scale length $h_R = 3~{\rm kpc}$ and scale height $h_z = 0.3~{\rm kpc}$.


To derive the circular velocity curve from the Jeans equation \ref{eq:vcgeneral}, we rewrite it as
\begin{equation}
\label{eq:vcfromjeans}
V_{\rm c}^2 =
\langle \varv_\Theta^2\rangle
-
\langle \varv_R^2\rangle
+
R
\left[
\frac{1}{\nu}\frac{\partial (\nu\langle \varv_R^2\rangle)}{\partial R}
+
\frac{1}{\nu}\frac{\partial (\nu\langle \varv_R v_z\rangle)}{\partial z}
\right].
\end{equation}
Then, rewriting this equation with gradients in $R$ and $Z$ in logarithmic form
\begin{align}
V_{\rm c}^2(R,z)
&=
\langle \varv_\Theta^2\rangle - \langle \varv_R^2\rangle
\nonumber\\
&\quad
+ R\left[
\frac{\partial \langle \varv_R^2\rangle}{\partial R}
+
\langle \varv_R^2\rangle\frac{\partial\ln\nu}{\partial R}
+
\frac{\partial \langle \varv_R \varv_z\rangle}{\partial z}
+
\langle \varv_R \varv_z\rangle\frac{\partial\ln\nu}{\partial z}
\right],
\label{eq:vcfinal}
\end{align}
and substituting into it the expression for $\nu(R, Z)$, we obtain the final expression for the circular velocity in a form convenient for calculation:

\begin{align}
V_{\rm c}^2(R,z)
& = \langle\varv_\Theta\rangle^2 + \sigma_\Theta^2 + (\langle\varv_R\rangle^2 + \sigma_R^2) \left(\frac{R-h_R}{h_R}\right) 
  - 2R\langle\varv_R\rangle\frac{\partial\langle\varv_R\rangle}{\partial R} \nonumber\\
& - R\frac{\partial\sigma_R^2}{\partial R} + \frac{R}{h_z}\frac{z}{|z|}\langle\varv_R\varv_Z\rangle - R\frac{\partial\langle\varv_R\varv_Z\rangle}{\partial z}
\end{align}
where $\langle \varv^2\rangle$ is the average value of the squared velocity for each component, which for a specific bin, i.e. a sphere with radius d = 1 kpc, can be written as:
\begin{equation}
\langle \varv^2\rangle = \langle \varv\rangle^2 + \sigma^2 + \sigma_{\rm obs}^2,
\end{equation}
where $\sigma^2$ is the intrinsic velocity dispersion in the bin,
$\langle \varv\rangle$ the mean velocity, and $\sigma_{\rm obs}^2$ the propagated
measurement variance transformed into the Galactocentric system. The second
term in equation~(\ref{eq:vcfinal}) is the asymmetric-drift correction, while
the third term represents the cross-term contribution. Equation
(\ref{eq:vcfinal}) requires no assumptions other than the tracer-density model
itself.

When transforming the observed mean azimuthal velocity
$\langle \varv_\Theta^2\rangle$ into the circular velocity
$\langle \varv_{\rm c}^2\rangle$, the quantities
$\langle \varv_R^2\rangle$, $\langle \varv_\Theta^2\rangle$,
$\langle \varv_Z^2\rangle$ and $\langle \varv_R \varv_Z\rangle$
determine the `anisotropy correction' of the velocity field, i.e. how
different the stellar velocity dispersions are in the radial, azimuthal and
vertical directions. In the cylindrical coordinates $(R,\Theta,Z)$, the anisotropy
is usually characterized by ratios of dispersions, which form the 
velocity ellipsoid parameters. By computing the second-order moments (dispersions
and covariances) of stellar velocities, we construct the full dispersion tensor,
which characterizes the spatial orientation and shape of the velocity
ellipsoid:
\begin{equation}
C_{\rm obs} =
\begin{pmatrix}
\sigma_{\varv_R}^2 &  \sigma_{\varv_R\varv_\Theta}  & \sigma_{\varv_R \varv_Z} \\
\sigma_{\varv_\Theta\varv_R} & \sigma_{\varv_\Theta}^2 & \sigma_{\varv_\Theta v_Z} \\
\sigma_{\varv_Z \varv_R} & \sigma_{\varv_Z \varv_\Theta} & \sigma_{\varv_Z}^2
\end{pmatrix}.
\label{eq:truecovmatrix}
\end{equation}

The described algorithm for computing dispersion-tensor components using a
spherical bin allows us to obtain, for each bin, a well-defined dispersion
value associated with the centroid coordinates $(R_i,\Theta_i)$. This sets us free
from the need to approximate them with analytic
functions. However, real velocity fields are non-stationary and non-uniform.
For example, $\varv_R$ inside a bin systematically depends on radius $R$ (e.g.
due to density waves or large-scale streaming flows). Within a bin, the radius
can vary by up to 2~kpc, which already results in systematic differences of
tens of km\,s$^{-1}$. Therefore, even after subtracting the mean velocity,
residual gradients are still present and contribute to the measured dispersions
that are not ``noise'' in the statistical sense. Thus, the measured dispersion
is a mixture:
\begin{equation}
\sigma^2_{\rm measured}
=
\sigma^2_{\rm true}
+
\sigma^2_{\rm gradient}
+
\sigma^2_{\rm stream}.
\end{equation}

For this reason, instead of simply averaging the velocities in each spherical
region, we determine all 12 kinematic parameters of the Ogorodnikov--Milne
model \citep{Fedorov2023}. The difference between the observed stellar
velocity and the model prediction (the O--M equation residuals) is treated as a
residual velocity whose distribution in each cell becomes close to Gaussian.
These residual velocities represent the peculiar stellar velocities affected
by measurement errors.

To compute the dispersions of these peculiar velocities, we use for each star
the measured values (with corrected parallax)
$(d, \mu_\alpha, \mu_\delta, \varv_r)$
and their uncertainties
$(\sigma_d, \sigma_{\mu_\alpha}, \sigma_{\mu_\delta}, \sigma_{\varv_r})$,
obtained from the \textit{Gaia} astrometry, and transform them into the required
coordinate system. If one simply takes the residual velocities and computes the
dispersion, the result contains both the true peculiar-velocity dispersion and
the measurement noise. Therefore, we compute the dispersions not directly but
using the covariance matrices of the propagated measurement errors.

For each star, we construct a $3\times3$ covariance matrix of the propagated
velocity uncertainties:
\begin{equation}
E_i =
\begin{pmatrix}
e^2_{\varv_R} & e_{\varv_R \varv_\Theta} & e_{\varv_R \varv_Z} \\
e_{\varv_\Theta \varv_R} & e^2_{\varv_\Theta} & e_{\varv_\Theta \varv_Z} \\
e_{\varv_Z \varv_R} & e_{\varv_Z \varv_\Theta} & e^2_{\varv_Z}
\end{pmatrix},
\label{eq:Ei_matrix}
\end{equation}
The elements of this matrix were obtained by linearly propagating the errors of the astrometric parameters using the Jacobian $J_i$ and $C_{{\rm astro},i}$ is the covariance matrix of the errors of the astrometric measurements from $Gaia$. Formally:
\begin{equation}
E_i = J_i\, C_{{\rm astro},i}\, J_i^{\rm T},
\end{equation}

All matrices $E_i$ in the bin are averaged 
and the `true' covariance tensor is obtained after subtracting the
contribution of the measurement noise:
\begin{equation}
C_{\rm true} = C_{\rm obs} - \langle E\rangle,
\end{equation}
yielding the corrected values of
$\sigma_{\varv_R}$, $\sigma_{\varv_\Theta}$, $\sigma_{\varv_Z}$ and $\sigma_{\varv_{R}\varv_Z}$.

\begin{figure*}
\centering
\resizebox{\hsize}{!}
   {\includegraphics{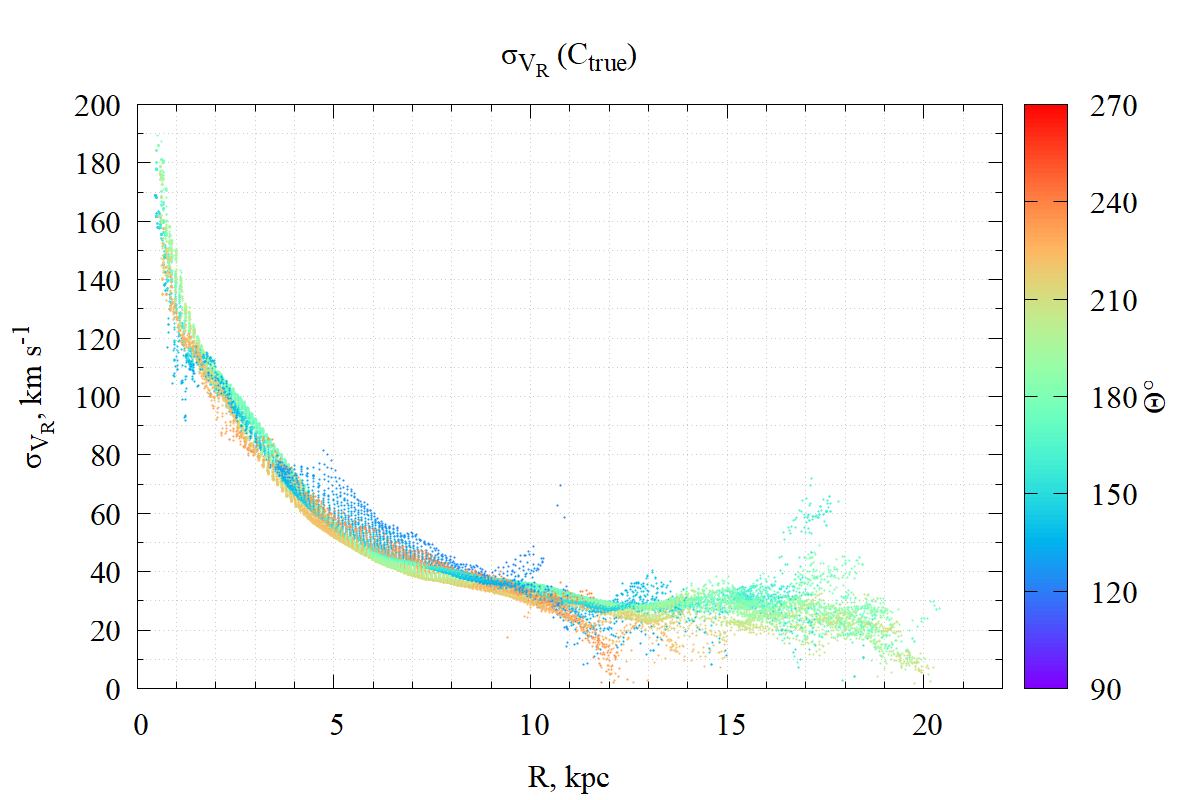}
    \includegraphics{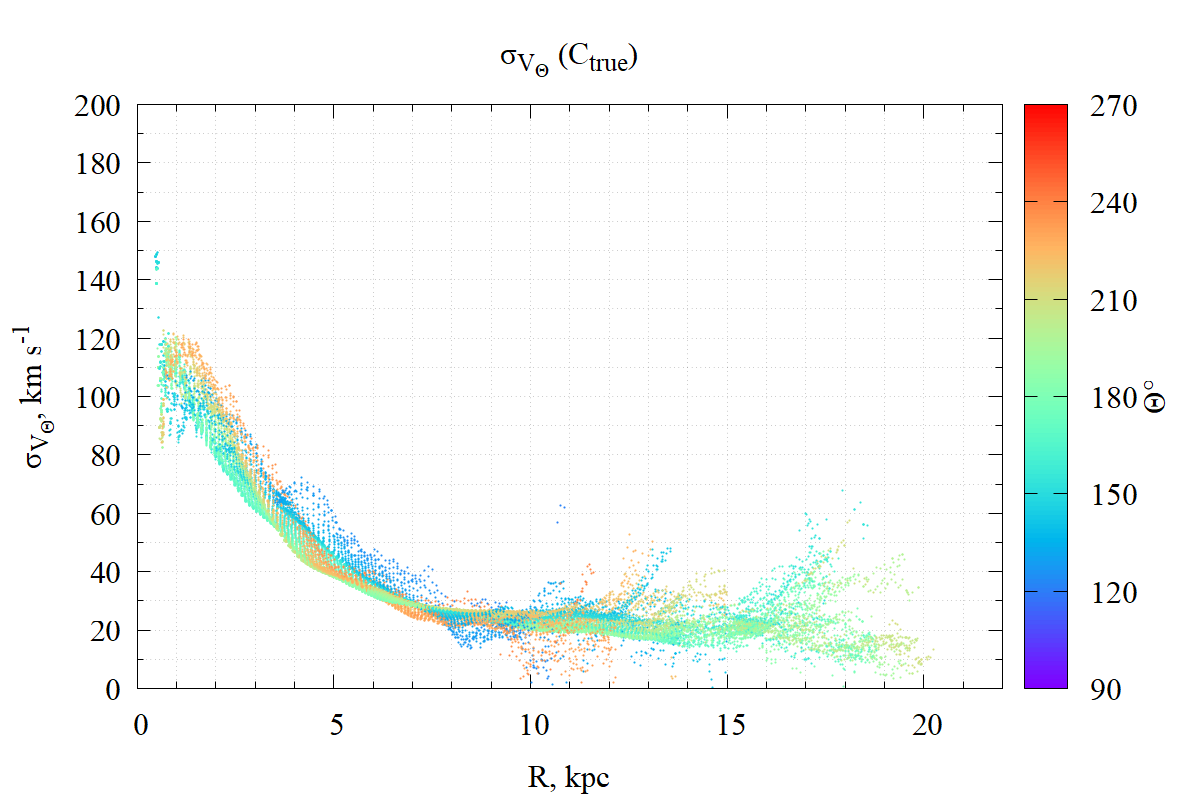}}
\resizebox{\hsize}{!}    
   {\includegraphics{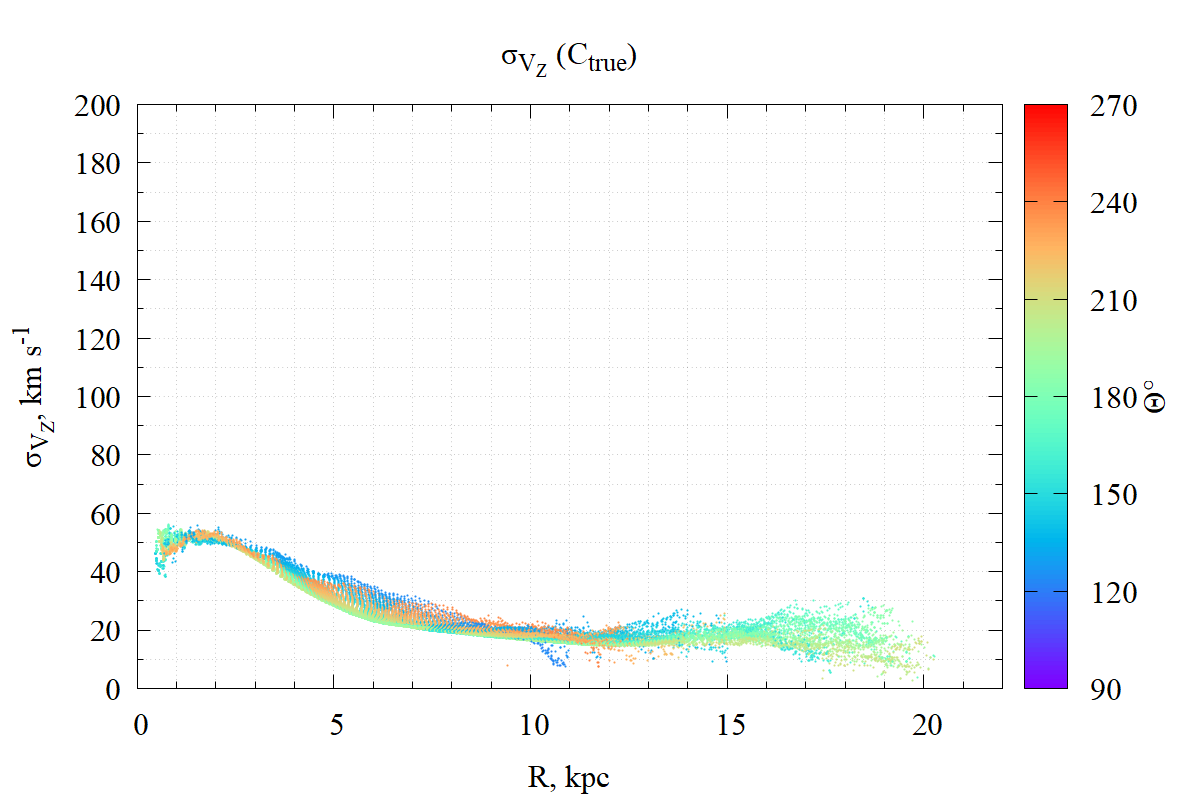}
    \includegraphics{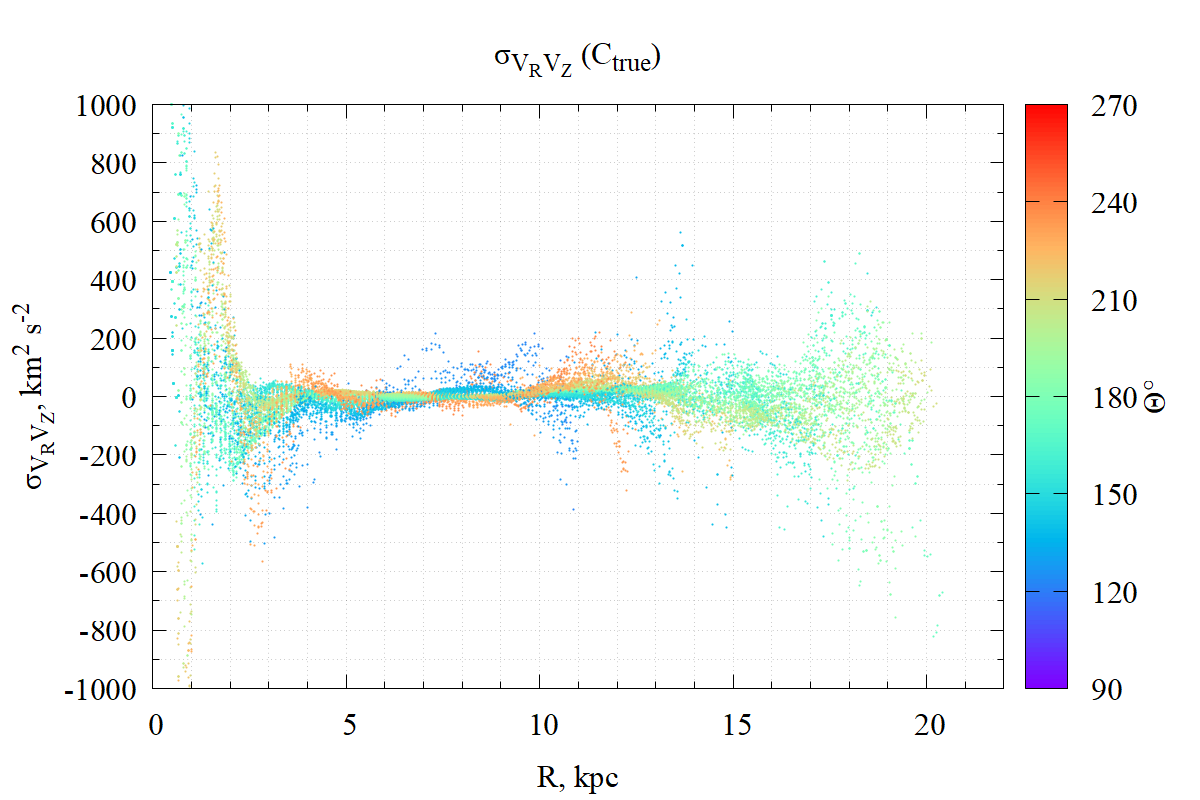}}
  \caption{$\sigma\varv_R, \sigma\varv_\Theta, \sigma\varv_Z$ and $\sigma\varv_R\varv_Z$ as functions of Galactocentric coordinates.}
\label{fig:sigmas}
\end{figure*}

In this way, we "cleanse" the dispersion tensor from the contribution of measurement errors and obtain the corresponding values ($\sigma\varv_R, \sigma\varv_\Theta, \sigma\varv_Z$, and $\sigma\varv_{RZ}$).
The figure \ref{fig:sigmas} shows ($\sigma\varv_R, \sigma\varv_\Theta, \sigma\varv_Z$, and $\sigma\varv_{RZ}$) obtained in the Galactic plane as functions of distance $R$. 

\begin{figure}
    \centering
    \includegraphics[width=1\linewidth]{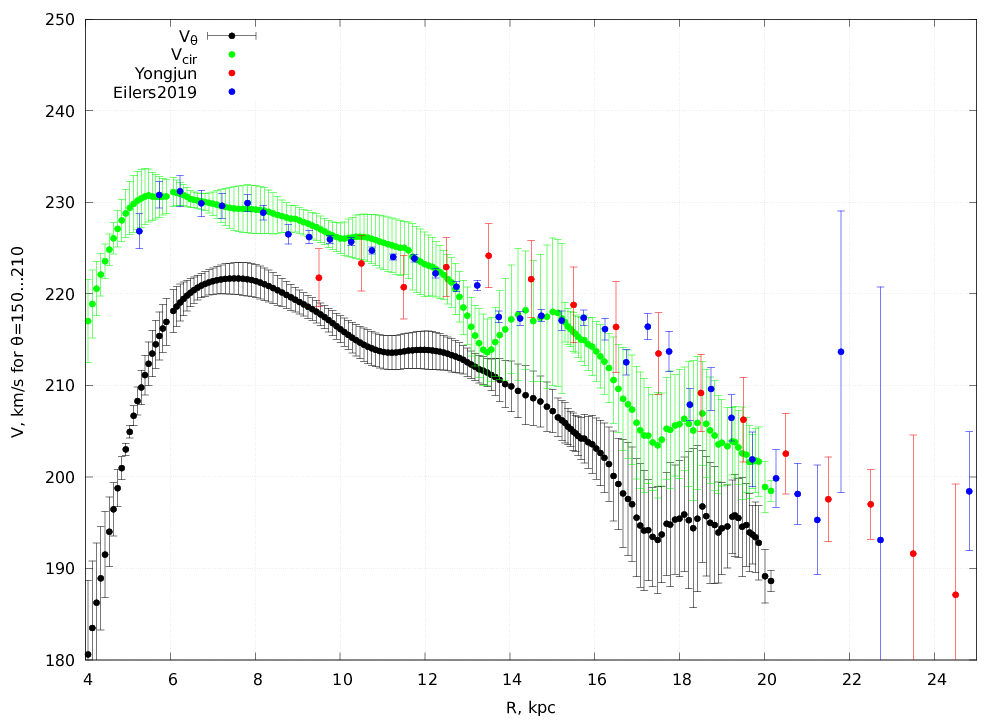}
    \caption{Averaged $V_{\rm c}$ and $V_\Theta$ as a function of the Galactocentric distance along the direction $\Theta = 180^\circ$, determined from stars contained in spheres with centers located at $Z=0.0$ kpc. Averaging of $V_{\rm c}$ and $V_\Theta$ was performed in the ranges of azimuthal angles $\pm30^\circ$ with respect to to $\Theta = 180^\circ$.}
    \label{fig:VthVcr_Z0}
\end{figure}

When computing $V_{\rm c}$ through equation \ref{eq:vcfinal}, derivatives of the form $\partial\langle\varv^2_R\rangle/\partial R$, $\partial\langle\varv_R\varv_Z\rangle/\partial Z$ appear, which need to be determined. Taking these derivatives directly from observational data yields a `jagged' dispersion profile due to noise, sample sparseness, and distance errors. To reduce noise, we did not approximate the dispersions with analytical functions, but instead calculated numerical derivatives based on dispersion values at points spaced 100 pc apart and contained within each sphere of radius $R = 1$kpc. Unfortunately, this did not radically reduce noise, but it did provide the opportunity to rely on real, unsmoothed data and eliminate the need for model representations.
The derived quantities were then substituted into the Jeans equation to obtain
the circular velocity $V_{\rm c}(R)$ in each spherical region of radius
1~kpc. The dependence of $V_{\rm c}$ on Galactocentric radius at
$\Theta = 180^\circ$ is shown in Fig.~\ref{fig:VthVcr_Z0}.

\begin{figure*}
\centering
\resizebox{\hsize}{!}
   {\includegraphics{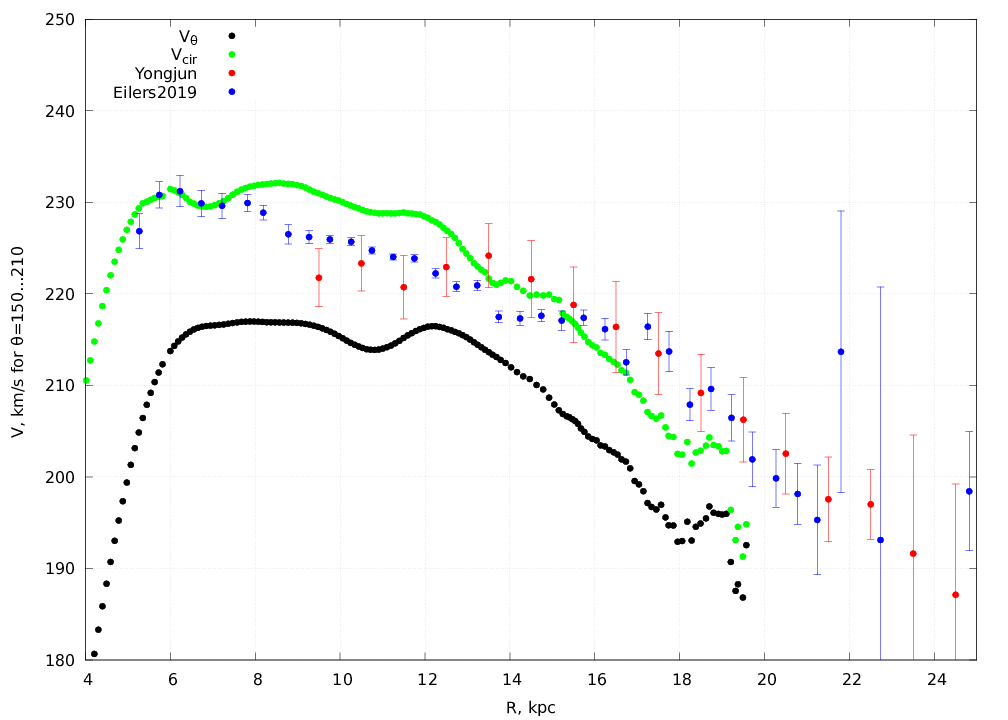}
    \includegraphics{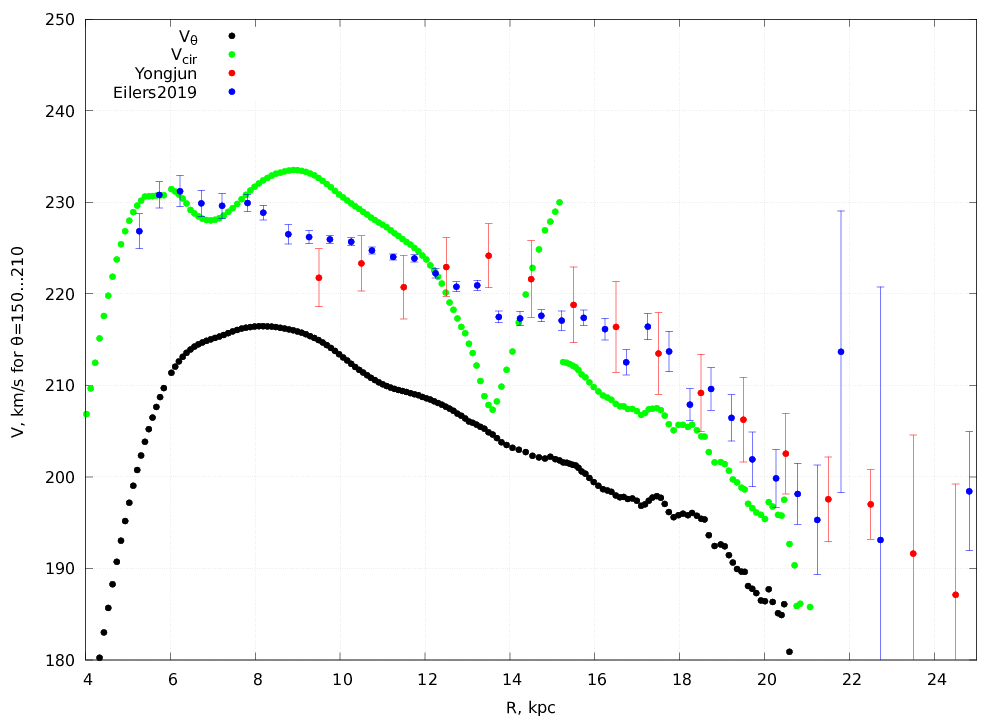}}
\resizebox{\hsize}{!}    
   {\includegraphics{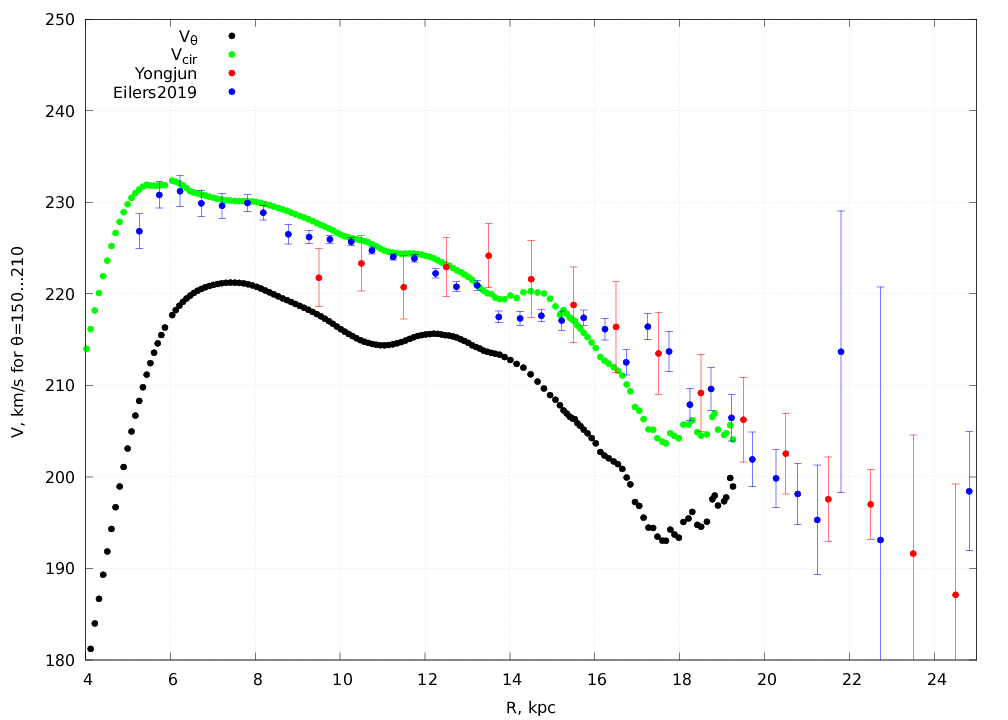}
    \includegraphics{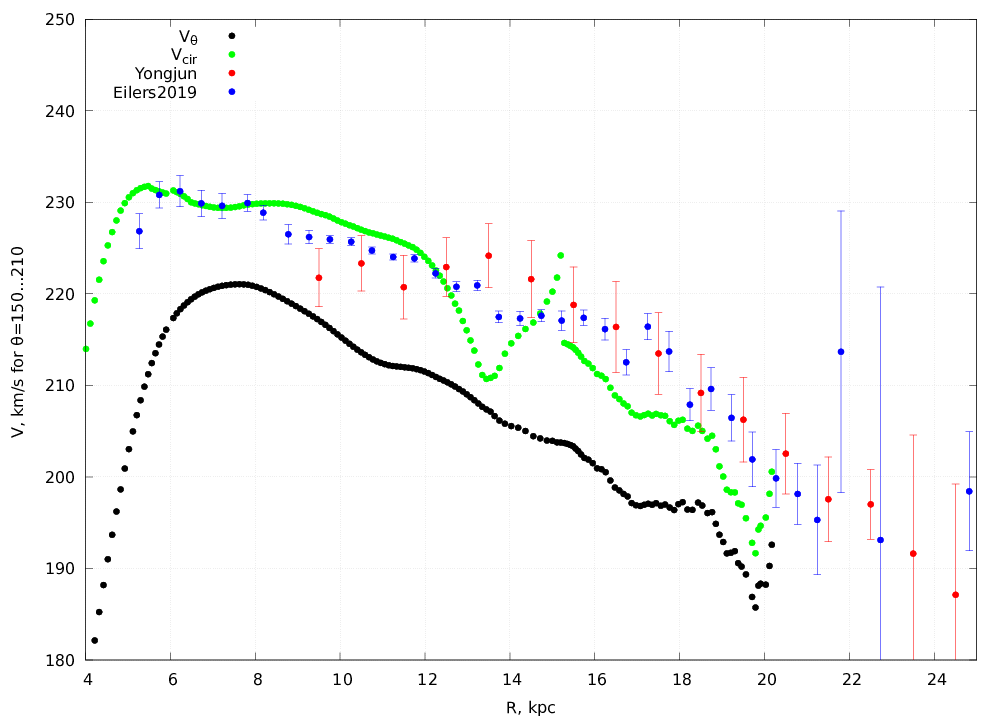}}
  \caption{Averaged $V_{\rm c}$ and $V_\Theta$ as a function of the Galactocentric distance along the direction $\Theta = 180^\circ$, determined from stars contained in spheres with centers located at $Z\pm1.0$ and $Z\pm0.5$ kpc. Averaging of $V_{\rm c}$ and $V_\Theta$ was performed in the ranges of azimuthal angles $\pm30^\circ$ with respect to to $\Theta = 180^\circ$.}
\label{fig:VthsVcrs}
\end{figure*}

Figure \ref{fig:VthsVcrs} shows $V_{\rm c}$ and $V_\Theta$ as a function of Galactocentric distance $R$. To make the comparison easier, the three panels display the functions corresponding to stars contained within spheres centered at Z = -1.0 and -0.5 kpc (left), Z = 0.0 kpc (middle) and Z = 1.0 and 0.5 kpc (right). Each point on the figure represents the value of $V_{\rm c}$ and $V_\Theta$  in specific spheres ($R=1$ kpc), separated by 100 pc. Averaging over $\Theta$ is performed in the range of $150^\circ-210^\circ$.


The rotation curve looks smoother than the circular velocity curve. However, despite some fluctuations in $V_{\rm c}$, the general trend, a decrease in velocity with distance, is clearly observed. At the same time, it is clearly visible that the curves differ somewhat in their slope, especially at large distances. The difference in amplitude is approximately $15-25$ km s$^{-1}$. According to Fig. \ref{fig:VthVcr_Z0}, the $V_{\rm c}$ and $V_\Theta$ curves are not linear functions in the distance range of $6-20$ kpc. Therefore, approximating their slope with a straight line seems inappropriate. Although in general (on average) the slopes of both curves are consistent, starting from about 16 kpc, we see that the overall slope of our curves is noticeably greater than that of the curve from \citet{Eilers2019}. We also note that the value of our circular velocity up to 15 kpc almost coincides with the values of \citet{Eilers2019}, although our curve is much more detailed.

\section{Circular velocity versus galactic azimuth and rotation curves in z-layers}
\label{sec:circvel}

\begin{figure*}
\centering
\resizebox{\hsize}{!}
   {\includegraphics{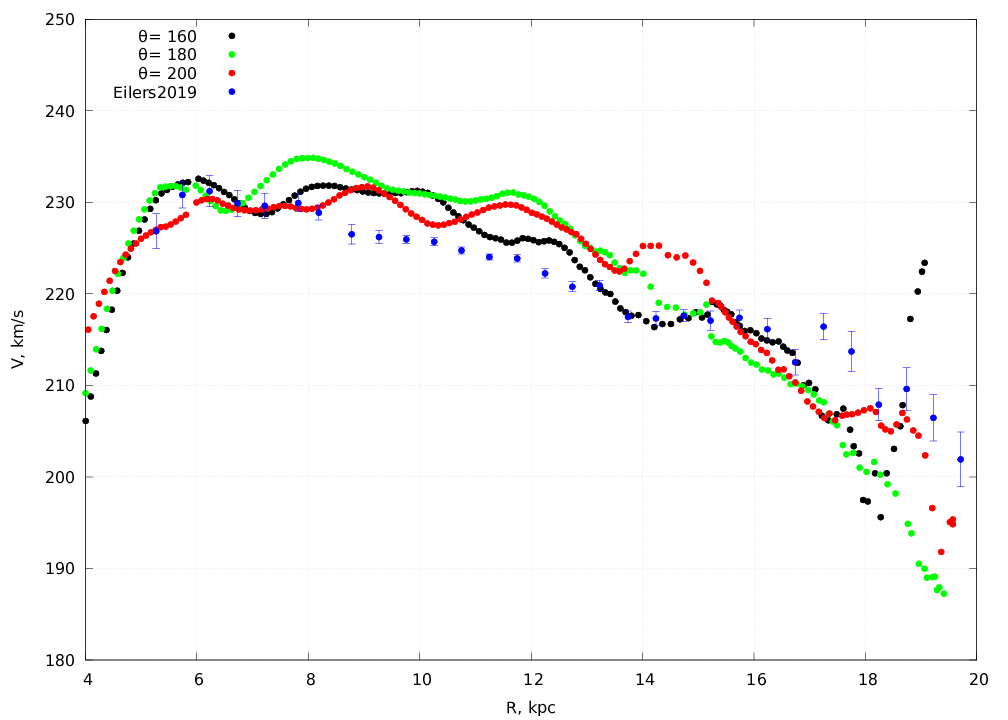}
    \includegraphics{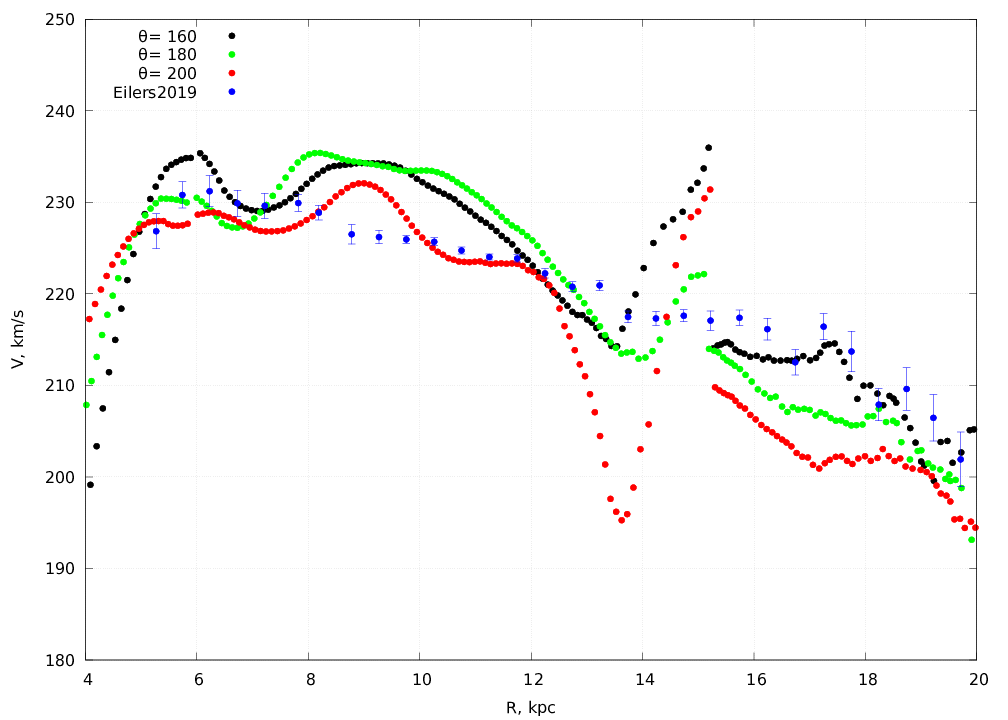}}
\resizebox{\hsize}{!}    
   {\includegraphics{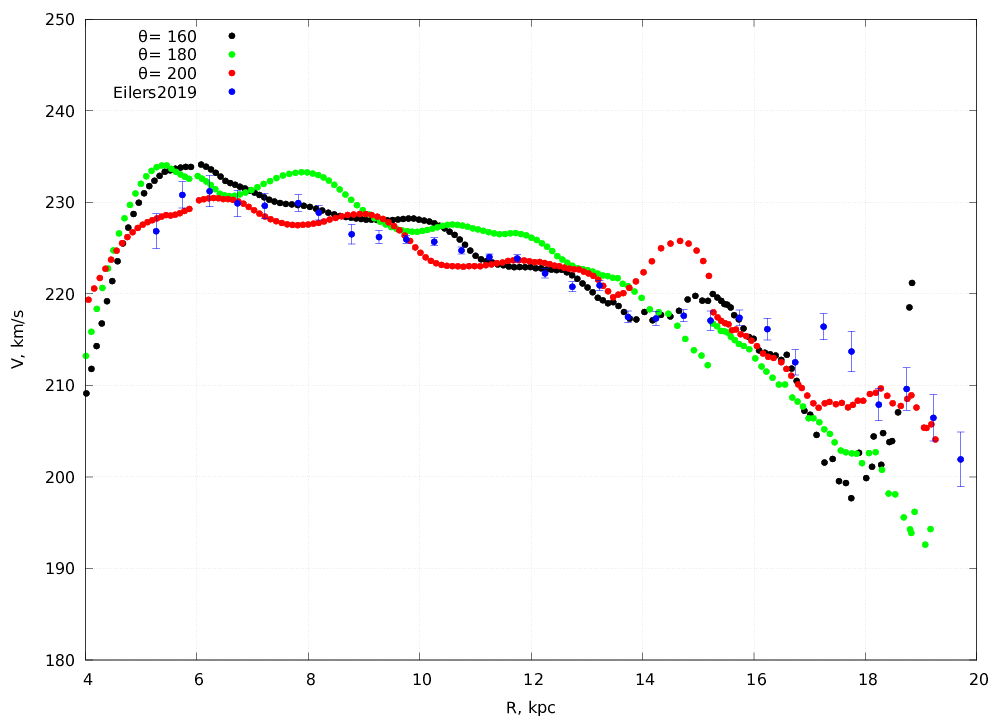}
    \includegraphics{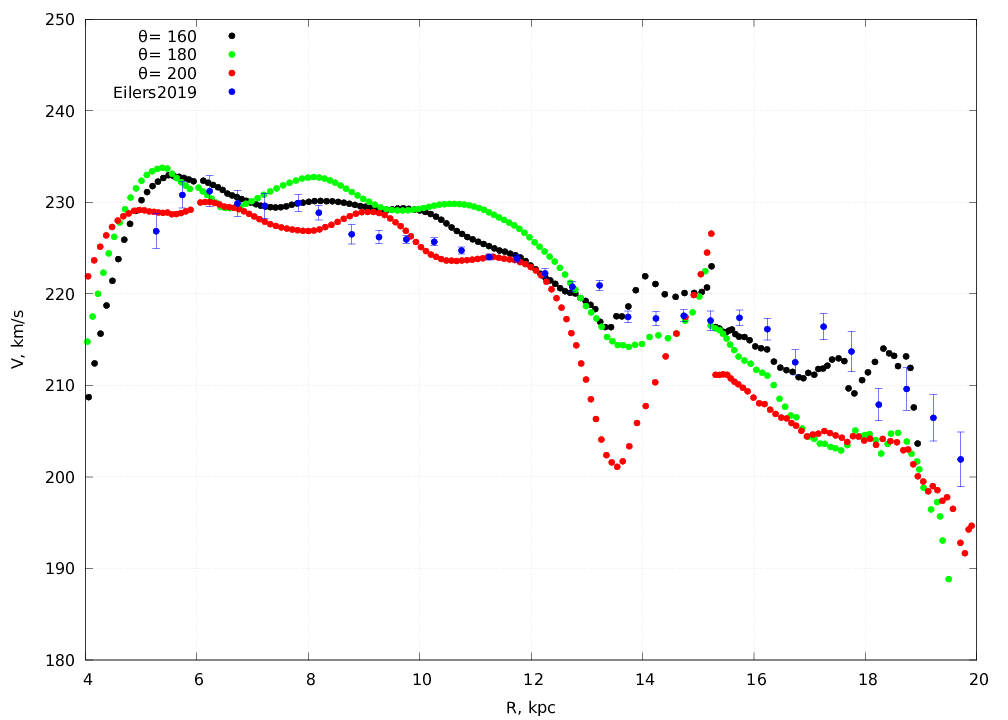}}
  \caption{The velocity $V$ as a function of Galactocentric distance along various azimuthal directions at heights $Z = \pm0.5$ and $\pm1.0$ kpc. The dependences were obtained for $\Theta = 160^\circ, 180^\circ$, and $200^\circ$ by averaging them over azimuthal angle ranges of $\pm10^\circ$ relative to each specified $\Theta$.}
\label{fig:Vcirs}
\end{figure*}

\begin{figure}
    \centering
    \includegraphics[width=1\linewidth]{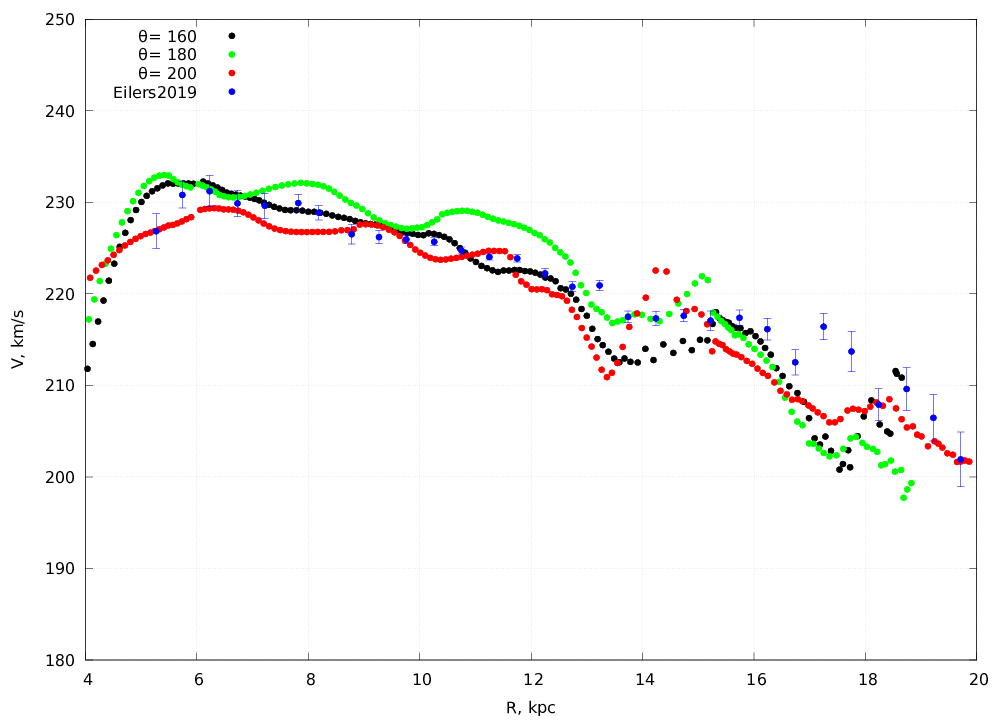}
    \caption{The velocity $V$ as a function of Galactocentric distance along various azimuthal directions at heights $Z=0$ kpc. The dependences were obtained for $\Theta = 160^\circ, 180^\circ$, and $200^\circ$ by averaging them over azimuthal angle ranges of $\pm10^\circ$ relative to each specified $\Theta$.}
    \label{fig:Vcir_Z0}
\end{figure}

We analyze the behavior of $V_{\rm c}$ and $V_\Theta$ over the range of $\Theta$ angles from 150$^\circ$ to 210$^\circ$. Figures \ref{fig:Vcirs}, \ref{fig:Vcir_Z0} show the velocities as a function of $R$ and $\Theta$. The plots are shown for negative heights on the left and for positive heights on the right.

\begin{figure}
\centering
\resizebox{\hsize}{!}
   {\includegraphics{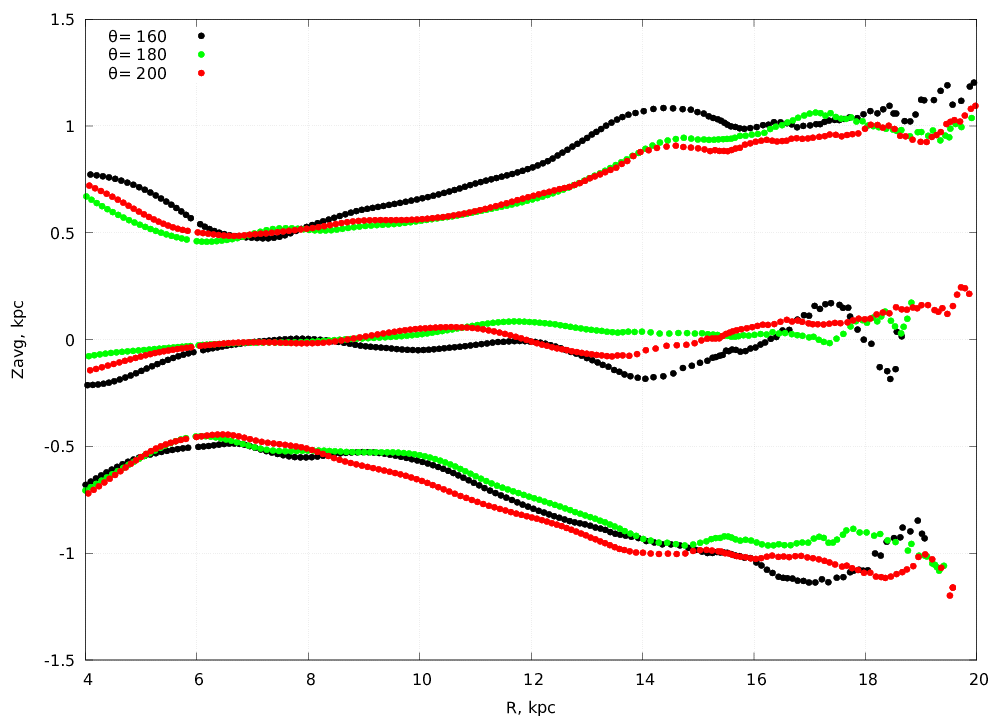} }
  \caption{The heights of the centroids as a function of Galactocentric distance with the centers of the spheres located at $Z=-1, Z=0$, and $Z=1$ kpc.}
\label{fig:Zavgs}
\end{figure}

The circular velocity is presented only in Fig. \ref{fig:Vcir_Z0}, since only the Galactic mid-plane is the plane of dynamical symmetry of the potential. Figs. \ref{fig:Vcirs} show the centroid velocities at various height layers. These velocities are not referred to the Galactic mid-plane, but to a certain `effective plane' by the heights of the centroids.

Figure. \ref{fig:Zavgs} shows the heights of the centroids as a function of distance for different $Z$. The tilts are caused by several physical factors that prevent us from interpreting $V_{\rm c}$ as a classical circular velocity. These `circular velocities' in the figure represent rotation curves for a population with average $Z$ and reflect the kinematics of a mixture of the thin and thick disks, and part of the halo at large $R$. Essentially, their behavior reflects the influence of the transition from disk to halo, the influence of the vertical structure of the potential and asymmetric drift as a function of $R$ and $Z$, as well as the influence of disk oscillations and warp. This is the physical meaning of the presented quantities, and it does not coincide at all with the physical meaning of the circular velocity for the thin disk. It is clearly seen from the figure that in the Galactic plane, no significant difference between the dependencies at different azimuths is observed up to $\sim$15 kpc. However, starting at $15-16$ kpc, the differences become large not only in the Galactic plane but also at various Z-heights, indicating a more complex velocity field in the ($R, z$) plane compared to the inner disk. Since most studies refer to the circular velocity curve from \citet{Eilers2019} and often make comparisons with it, we also show this function in blue in all figures.

It is clearly seen that $V_{\rm c}$, obtained by averaging in the range of azimuthal angles $\pm30^\circ$, is significantly closer to the blue dots \citep[the curve from][]{Eilers2019}. This likely indicates that the curve in \citet{Eilers2019} was obtained by averaging over all azimuths available.

\section{Summary and Conclusions}
\label{sec:conclusions}

First, we note that the results obtained do not contradict other published findings \citep[e.g., ][and others]{Eilers2019, Ou2024, Wang2023, Koop2024}. At the same time, it should be emphasized that they were derived from an RGB sample (4,547,980 stars) taken solely from $Gaia$ DR3, with minimal reliance on model-based data. In addition, the circular and rotational velocity curves are highly detailed, as they were constructed using spatial nodes separated by only 100 pc.
The resulting rotation curve and circular velocity curve exhibit some features not always observed in other studies. These features are present at distances of $\sim13$ and $\sim18$ kpc. However, they can sometimes be found in other studies, albeit in a significantly smoother form \citep[see, for example, ][]{Koop2024}. To verify whether the dip at 18 kpc is real, we constructed a rotation curve for our sample using the photogeometric distances obtained by \citet{Bailer-Jones2021} as a distance estimate.



A noticeable velocity decrease turned out to be also present at radii of $\sim13-18$ kpc. Some discrepancy in the specific distance values where the velocity minimum is reached is apparently due to the use of different distance estimates. Therefore, the noticeable decrease in rotation velocity near these points is, in our opinion, real and is not determined by parallax inaccuracies, but rather indicates the presence of a rather complex velocity field at this radius.

When approximating the derived circular velocity with a linear function, we see a steady decrease in the circular velocity by $2.3\pm0.5 $ km s$^{-1}$ kpc$^{-1}$ over the entire range of distances used, which can be represented as:
\begin{equation}
   V_{\rm c} (R_\odot)=(229.63\pm0.30)\,{\rm km\,s}^{-1} - (2.29\pm0.05)\,{\rm km\,s^{-1}\,kpc}^{-1}\cdot(R-R_\odot) \nonumber
\end{equation}
This is consistent with previous studies by \citet{Wang2023, Zhou2023, Ou2024} and \citet{Jiao2023}.

We observe rotation curves that are not quite identical at different $\Theta$ angles, indicating a dependence of the circular velocity on azimuth. Although the differences are small, they do exist and are particularly noticeable at large $\Theta$ angles, where they can reach $\sim$20 km s$^{-1}$ and be completely different from the curve at $\Theta=180$. In this sense, we regard a given circular velocity curve as a local characteristic, the application of which to the entire Galaxy seems unjustified to us. The lack of data for a significantly wider range of $\Theta$ angles and the limited distance range require great caution in determining the shape of the $V_{\rm c}$ curve and its derivative $dV_{\rm c}/dR$, especially for determining the Galactic mass. However, the use of detailed $V_{\rm c}$ and $V_\Theta$, along with additional indicators, e.g. morphology, composition, etc., may prove effective characteristics for use as indicators of Milky Way analogues.



\section{Acknowledgements}
\label{sec:acknowledgements}

This work has made use of data from the European Space Agency (ESA) mission {\it Gaia} (\url{https://www.cosmos.esa.int/gaia}), processed by the {\it Gaia} Data Processing and Analysis Consortium (DPAC, \url{https://www.cosmos.esa.int/web/gaia/dpac/consortium}). Funding for the DPAC has been provided by national institutions, in particular the institutions participating in the {\it Gaia} Multilateral Agreement.

This work is supported by the National Research Foundation of Ukraine, Project No. 2023.03/0188 \citep[][]{Vavilova2024}, and the Ministry of Education and Science of Ukraine.


\section*{Data availability}
\addcontentsline{toc}{section}{Data availability}

The used catalogue data are available in a standardized format for readers via the CDS (https://cds.u-strasbg.fr). 
The software code used in this paper can be made available on personal request by e-mail: \href{mailto:akhmetovvs@gmail.com} {akhmetovvs@gmail.com}.

\bibliographystyle{mnras}
\bibliography{references}

@ARTICLE{Abuter2021,
       author = {{GRAVITY Collaboration} and {Abuter}, R. and {Amorim}, A. and {Baub{\"o}ck}, M. and {Berger}, J.~P. and {Bonnet}, H. and {Brandner}, W. and {Cl{\'e}net}, Y. and {Davies}, R. and {de Zeeuw}, P.~T. and {Dexter}, J. and {Dallilar}, Y. and {Drescher}, A. and {Eckart}, A. and {Eisenhauer}, F. and {F{\"o}rster Schreiber}, N.~M. and {Garcia}, P. and {Gao}, F. and {Gendron}, E. and {Genzel}, R. and {Gillessen}, S. and {Habibi}, M. and {Haubois}, X. and {Hei{\ss}el}, G. and {Henning}, T. and {Hippler}, S. and {Horrobin}, M. and {Jim{\'e}nez-Rosales}, A. and {Jochum}, L. and {Jocou}, L. and {Kaufer}, A. and {Kervella}, P. and {Lacour}, S. and {Lapeyr{\`e}re}, V. and {Le Bouquin}, J. -B. and {L{\'e}na}, P. and {Lutz}, D. and {Nowak}, M. and {Ott}, T. and {Paumard}, T. and {Perraut}, K. and {Perrin}, G. and {Pfuhl}, O. and {Rabien}, S. and {Rodr{\'\i}guez-Coira}, G. and {Shangguan}, J. and {Shimizu}, T. and {Scheithauer}, S. and {Stadler}, J. and {Straub}, O. and {Straubmeier}, C. and {Sturm}, E. and {Tacconi}, L.~J. and {Vincent}, F. and {von Fellenberg}, S. and {Waisberg}, I. and {Widmann}, F. and {Wieprecht}, E. and {Wiezorrek}, E. and {Woillez}, J. and {Yazici}, S. and {Young}, A. and {Zins}, G.},
        title = "{Improved GRAVITY astrometric accuracy from modeling optical aberrations}",
      journal = {\aap},
         year = 2021,
        month = mar,
       volume = {647},
          eid = {A59},
        pages = {A59},
          doi = {10.1051/0004-6361/202040208},
archivePrefix = {arXiv},
       eprint = {2101.12098},
 primaryClass = {astro-ph.GA},
       adsurl = {https://ui.adsabs.harvard.edu/abs/2021A&A...647A..59G}
}

@ARTICLE{Akhmetov2024,
    author = {{Akhmetov}, V.~S. and {Bucciarelli}, B. and {Crosta}, M. and {Lattanzi}, M.~G. and {Spagna}, A. and {Re Fiorentin}, P. and {Bannikova}, E. Yu},
        title = "{A new kinematic model of the Galaxy: analysis of the stellar velocity field from Gaia Data Release 3}",
      journal = {\mnras},
         year = 2024,
        month = may,
       volume = {530},
       number = {1},
        pages = {710-729},
          doi = {10.1093/mnras/stae772},
archivePrefix = {arXiv},
       eprint = {2307.08527},
 primaryClass = {astro-ph.GA},
       adsurl = {https://ui.adsabs.harvard.edu/abs/2024MNRAS.530..710A}
}

@ARTICLE{Andrae2023,
       author = {{Andrae}, R. and {Fouesneau}, M. and {Sordo}, R. and {Bailer-Jones}, C.~A.~L. and {Dharmawardena}, T.~E. and {Rybizki}, J. and {De Angeli}, F. and {Lindstr{\o}m}, H.~E.~P. and {Marshall}, D.~J. and {Drimmel}, R. and {Korn}, A.~J. and {Soubiran}, C. and {Brouillet}, N. and {Casamiquela}, L. and {Rix}, H.-W. and {Abreu Aramburu}, A. and {{\'A}lvarez}, M.~A. and {Bakker}, J. and {Bellas-Velidis}, I. and {Bijaoui}, A. and {Brugaletta}, E. and {Burlacu}, A. and {Carballo}, R. and {Chaoul}, L. and {Chiavassa}, A. and {Contursi}, G. and {Cooper}, W.~J. and {Creevey}, O.~L. and {Dafonte}, C. and {Dapergolas}, A. and {de Laverny}, P. and {Delchambre}, L. and {Demouchy}, C. and {Edvardsson}, B. and {Fr{\'e}mat}, Y. and {Garabato}, D. and {Garc{\'\i}a-Lario}, P. and {Garc{\'\i}a-Torres}, M. and {Gavel}, A. and {Gomez}, A. and {Gonz{\'a}lez-Santamar{\'\i}a}, I. and {Hatzidimitriou}, D. and {Heiter}, U. and {Jean-Antoine Piccolo}, A. and {Kontizas}, M. and {Kordopatis}, G. and {Lanzafame}, A.~C. and {Lebreton}, Y. and {Licata}, E.~L. and {Livanou}, E. and {Lobel}, A. and {Lorca}, A. and {Magdaleno Romeo}, A. and {Manteiga}, M. and {Marocco}, F. and {Mary}, N. and {Nicolas}, C. and {Ordenovic}, C. and {Pailler}, F. and {Palicio}, P.~A. and {Pallas-Quintela}, L. and {Panem}, C. and {Pichon}, B. and {Poggio}, E. and {Recio-Blanco}, A. and {Riclet}, F. and {Robin}, C. and {Santove{\~n}a}, R. and {Sarro}, L.~M. and {Schultheis}, M.~S. and {Segol}, M. and {Silvelo}, A. and {Slezak}, I. and {Smart}, R.~L. and {S{\"u}veges}, M. and {Th{\'e}venin}, F. and {Torralba Elipe}, G. and {Ulla}, A. and {Utrilla}, E. and {Vallenari}, A. and {van Dillen}, E. and {Zhao}, H. and {Zorec}, J.},
        title = "{Gaia Data Release 3. Analysis of the Gaia BP/RP spectra using the General Stellar Parameterizer from Photometry}",
      journal = {\aap},
         year = 2023,
        month = jun,
       volume = {674},
          eid = {A27},
        pages = {A27},
          doi = {10.1051/0004-6361/202243462},
archivePrefix = {arXiv},
       eprint = {2206.06138},
 primaryClass = {astro-ph.SR},
       adsurl = {https://ui.adsabs.harvard.edu/abs/2023A&A...674A..27A}
}

@ARTICLE{Antoja2018,
       author = {{Antoja}, T. and {Helmi}, A. and {Romero-G{\'o}mez}, M. and {Katz}, D. and {Babusiaux}, C. and {Drimmel}, R. and {Evans}, D.~W. and {Figueras}, F. and {Poggio}, E. and {Reyl{\'e}}, C. and {Robin}, A.~C. and {Seabroke}, G. and {Soubiran}, C.},
        title = "{A dynamically young and perturbed Milky Way disk}",
      journal = {\nat},
         year = 2018,
        month = sep,
       volume = {561},
       number = {7723},
        pages = {360-362},
          doi = {10.1038/s41586-018-0510-7},
archivePrefix = {arXiv},
       eprint = {1804.10196},
 primaryClass = {astro-ph.GA},
       adsurl = {https://ui.adsabs.harvard.edu/abs/2018Natur.561..360A}
}

@ARTICLE{Antoja2021,
       author = {{Gaia Collaboration} and {Antoja}, T. and {McMillan}, P.~J. and {Kordopatis}, G. and {Ramos}, P. and {Helmi}, A. and {Balbinot}, E. and {Cantat-Gaudin}, T. and {Chemin}, L. and {Figueras}, F. and {Jordi}, C. and {Khanna}, S. and {Romero-G{\'o}mez}, M. and {Seabroke}, G.~M. and {Brown}, A.~G.~A. and {Vallenari}, A. and {Prusti}, T. and {de Bruijne}, J.~H.~J. and {Babusiaux}, C. and {Biermann}, M. and {Creevey}, O.~L. and {Evans}, D.~W. and {Eyer}, L. and {Hutton}, A. and {Jansen}, F. and {Klioner}, S.~A. and {Lammers}, U. and {Lindegren}, L. and {Luri}, X. and {Mignard}, F. and {Panem}, C. and {Pourbaix}, D. and {Randich}, S. and {Sartoretti}, P. and {Soubiran}, C. and {Walton}, N.~A. and {Arenou}, F. and {Bailer-Jones}, C.~A.~L. and {Bastian}, U. and {Cropper}, M. and {Drimmel}, R. and {Katz}, D. and {Lattanzi}, M.~G. and {van Leeuwen}, F. and {Bakker}, J. and {Casta{\~n}eda}, J. and {De Angeli}, F. and {Ducourant}, C. and {Fabricius}, C. and {Fouesneau}, M. and {Fr{\'e}mat}, Y. and {Guerra}, R. and {Guerrier}, A. and {Guiraud}, J. and {Jean-Antoine Piccolo}, A. and {Masana}, E. and {Messineo}, R. and {Mowlavi}, N. and {Nicolas}, C. and {Nienartowicz}, K. and {Pailler}, F. and {Panuzzo}, P. and {Riclet}, F. and {Roux}, W. and {Sordo}, R. and {Tanga}, P. and {Th{\'e}venin}, F. and {Gracia-Abril}, G. and {Portell}, J. and {Teyssier}, D. and {Altmann}, M. and {Andrae}, R. and {Bellas-Velidis}, I. and {Benson}, K. and {Berthier}, J. and {Blomme}, R. and {Brugaletta}, E. and {Burgess}, P.~W. and {Busso}, G. and {Carry}, B. and {Cellino}, A. and {Cheek}, N. and {Clementini}, G. and {Damerdji}, Y. and {Davidson}, M. and {Delchambre}, L. and {Dell'Oro}, A. and {Fern{\'a}ndez-Hern{\'a}ndez}, J. and {Galluccio}, L. and {Garc{\'\i}a-Lario}, P. and {Garcia-Reinaldos}, M. and {Gonz{\'a}lez-N{\'u}{\~n}ez}, J. and {Gosset}, E. and {Haigron}, R. and {Halbwachs}, J. -L. and {Hambly}, N.~C. and {Harrison}, D.~L. and {Hatzidimitriou}, D. and {Heiter}, U. and {Hern{\'a}ndez}, J. and {Hestroffer}, D. and {Hodgkin}, S.~T. and {Holl}, B. and {Jan{\ss}en}, K. and {Jevardat de Fombelle}, G. and {Jordan}, S. and {Krone-Martins}, A. and {Lanzafame}, A.~C. and {L{\"o}ffler}, W. and {Lorca}, A. and {Manteiga}, M. and {Marchal}, O. and {Marrese}, P.~M. and {Moitinho}, A. and {Mora}, A. and {Muinonen}, K. and {Osborne}, P. and {Pancino}, E. and {Pauwels}, T. and {Recio-Blanco}, A. and {Richards}, P.~J. and {Riello}, M. and {Rimoldini}, L. and {Robin}, A.~C. and {Roegiers}, T. and {Rybizki}, J. and {Sarro}, L.~M. and {Siopis}, C. and {Smith}, M. and {Sozzetti}, A. and {Ulla}, A. and {Utrilla}, E. and {van Leeuwen}, M. and {van Reeven}, W. and {Abbas}, U. and {Abreu Aramburu}, A. and {Accart}, S. and {Aerts}, C. and {Aguado}, J.~J. and {Ajaj}, M. and {Altavilla}, G. and {{\'A}lvarez}, M.~A. and {{\'A}lvarez Cid-Fuentes}, J. and {Alves}, J. and {Anderson}, R.~I. and {Varela}, E. Anglada and {Audard}, M. and {Baines}, D. and {Baker}, S.~G. and {Balaguer-N{\'u}{\~n}ez}, L. and {Balog}, Z. and {Barache}, C. and {Barbato}, D. and {Barros}, M. and {Barstow}, M.~A. and {Bartolom{\'e}}, S. and {Bassilana}, J. -L. and {Bauchet}, N. and {Baudesson-Stella}, A. and {Becciani}, U. and {Bellazzini}, M. and {Bernet}, M. and {Bertone}, S. and {Bianchi}, L. and {Blanco-Cuaresma}, S. and {Boch}, T. and {Bombrun}, A. and {Bossini}, D. and {Bouquillon}, S. and {Bragaglia}, A. and {Bramante}, L. and {Breedt}, E. and {Bressan}, A. and {Brouillet}, N. and {Bucciarelli}, B. and {Burlacu}, A. and {Busonero}, D. and {Butkevich}, A.~G. and {Buzzi}, R. and {Caffau}, E. and {Cancelliere}, R. and {C{\'a}novas}, H. and {Carballo}, R. and {Carlucci}, T. and {Carnerero}, M.~I. and {Carrasco}, J.~M. and {Casamiquela}, L. and {Castellani}, M. and {Castro-Ginard}, A. and {Castro Sampol}, P. and {Chaoul}, L. and {Charlot}, P. and {Chiavassa}, A. and {Cioni}, M. -R.~L. and {Comoretto}, G. and {Cooper}, W.~J. and {Cornez}, T. and {Cowell}, S. and {Crifo}, F. and {Crosta}, M. and {Crowley}, C. and {Dafonte}, C. and {Dapergolas}, A. and {David}, M. and {David}, P. and {de Laverny}, P. and {De Luise}, F. and {De March}, R. and {De Ridder}, J. and {de Souza}, R. and {de Teodoro}, P. and {de Torres}, A. and {del Peloso}, E.~F. and {del Pozo}, E. and {Delgado}, A. and {Delgado}, H.~E. and {Delisle}, J. -B. and {Di Matteo}, P. and {Diakite}, S. and {Diener}, C. and {Distefano}, E. and {Dolding}, C. and {Eappachen}, D. and {Enke}, H. and {Esquej}, P. and {Fabre}, C. and {Fabrizio}, M. and {Faigler}, S. and {Fedorets}, G. and {Fernique}, P. and {Fienga}, A. and {Fouron}, C. and {Fragkoudi}, F. and {Fraile}, E. and {Franke}, F. and {Gai}, M. and {Garabato}, D. and {Garcia-Gutierrez}, A. and {Garc{\'\i}a-Torres}, M. and {Garofalo}, A. and {Gavras}, P. and {Gerlach}, E. and {Geyer}, R. and {Giacobbe}, P. and {Gilmore}, G. and {Girona}, S. and {Giuffrida}, G. and {Gomez}, A. and {Gonzalez-Santamaria}, I. and {Gonz{\'a}lez-Vidal}, J.~J. and {Granvik}, M. and {Guti{\'e}rrez-S{\'a}nchez}, R. and {Guy}, L.~P. and {Hauser}, M. and {Haywood}, M. and {Hidalgo}, S.~L. and {Hilger}, T. and {H{\l}adczuk}, N. and {Hobbs}, D. and {Holland}, G. and {Huckle}, H.~E. and {Jasniewicz}, G. and {Jonker}, P.~G. and {Juaristi Campillo}, J. and {Julbe}, F. and {Karbevska}, L. and {Kervella}, P. and {Kochoska}, A. and {Kontizas}, M. and {Korn}, A.~J. and {Kostrzewa-Rutkowska}, Z. and {Kruszy{\'n}ska}, K. and {Lambert}, S. and {Lanza}, A.~F. and {Lasne}, Y. and {Le Campion}, J. -F. and {Le Fustec}, Y. and {Lebreton}, Y. and {Lebzelter}, T. and {Leccia}, S. and {Leclerc}, N. and {Lecoeur-Taibi}, I. and {Liao}, S. and {Licata}, E. and {Lindstr{\o}m}, H.~E.~P. and {Lister}, T.~A. and {Livanou}, E. and {Lobel}, A. and {Madrero Pardo}, P. and {Managau}, S. and {Mann}, R.~G. and {Marchant}, J.~M. and {Marconi}, M. and {Marcos Santos}, M.~M.~S. and {Marinoni}, S. and {Marocco}, F. and {Marshall}, D.~J. and {Martin Polo}, L. and {Mart{\'\i}n-Fleitas}, J.~M. and {Masip}, A. and {Massari}, D. and {Mastrobuono-Battisti}, A. and {Mazeh}, T. and {Messina}, S. and {Michalik}, D. and {Millar}, N.~R. and {Mints}, A. and {Molina}, D. and {Molinaro}, R. and {Moln{\'a}r}, L. and {Montegriffo}, P. and {Mor}, R. and {Morbidelli}, R. and {Morel}, T. and {Morris}, D. and {Mulone}, A.~F. and {Munoz}, D. and {Muraveva}, T. and {Murphy}, C.~P. and {Musella}, I. and {Noval}, L. and {Ord{\'e}novic}, C. and {Orr{\`u}}, G. and {Osinde}, J. and {Pagani}, C. and {Pagano}, I. and {Palaversa}, L. and {Palicio}, P.~A. and {Panahi}, A. and {Pawlak}, M. and {Pe{\~n}alosa Esteller}, X. and {Penttil{\"a}}, A. and {Piersimoni}, A.~M. and {Pineau}, F. -X. and {Plachy}, E. and {Plum}, G. and {Poggio}, E. and {Poretti}, E. and {Poujoulet}, E. and {Pr{\v{s}}a}, A. and {Pulone}, L. and {Racero}, E. and {Ragaini}, S. and {Rainer}, M. and {Raiteri}, C.~M. and {Rambaux}, N. and {Ramos-Lerate}, M. and {Re Fiorentin}, P. and {Regibo}, S. and {Reyl{\'e}}, C. and {Ripepi}, V. and {Riva}, A. and {Rixon}, G. and {Robichon}, N. and {Robin}, C. and {Roelens}, M. and {Rohrbasser}, L. and {Rowell}, N. and {Royer}, F. and {Rybicki}, K.~A. and {Sadowski}, G. and {Sagrist{\`a} Sell{\'e}s}, A. and {Sahlmann}, J. and {Salgado}, J. and {Salguero}, E. and {Samaras}, N. and {Sanchez Gimenez}, V. and {Sanna}, N. and {Santove{\~n}a}, R. and {Sarasso}, M. and {Schultheis}, M. and {Sciacca}, E. and {Segol}, M. and {Segovia}, J.~C. and {S{\'e}gransan}, D. and {Semeux}, D. and {Siddiqui}, H.~I. and {Siebert}, A. and {Siltala}, L. and {Slezak}, E. and {Smart}, R.~L. and {Solano}, E. and {Solitro}, F. and {Souami}, D. and {Souchay}, J. and {Spagna}, A. and {Spoto}, F. and {Steele}, I.~A. and {Steidelm{\"u}ller}, H. and {Stephenson}, C.~A. and {S{\"u}veges}, M. and {Szabados}, L. and {Szegedi-Elek}, E. and {Taris}, F. and {Tauran}, G. and {Taylor}, M.~B. and {Teixeira}, R. and {Thuillot}, W. and {Tonello}, N. and {Torra}, F. and {Torra}, J. and {Turon}, C. and {Unger}, N. and {Vaillant}, M. and {van Dillen}, E. and {Vanel}, O. and {Vecchiato}, A. and {Viala}, Y. and {Vicente}, D. and {Voutsinas}, S. and {Weiler}, M. and {Wevers}, T. and {Wyrzykowski}, {\L}. and {Yoldas}, A. and {Yvard}, P. and {Zhao}, H. and {Zorec}, J. and {Zucker}, S. and {Zurbach}, C. and {Zwitter}, T.},
        title = "{Gaia Early Data Release 3. The Galactic anticentre}",
      journal = {\aap},
         year = 2021,
        month = may,
       volume = {649},
          eid = {A8},
        pages = {A8},
          doi = {10.1051/0004-6361/202039714},
archivePrefix = {arXiv},
       eprint = {2101.05811},
 primaryClass = {astro-ph.GA},
       adsurl = {https://ui.adsabs.harvard.edu/abs/2021A&A...649A...8G}
}

@ARTICLE{Bailer-Jones2021,
       author = {{Bailer-Jones}, C.~A.~L. and {Rybizki}, J. and {Fouesneau}, M. and {Demleitner}, M. and {Andrae}, R.},
        title = "{Estimating Distances from Parallaxes. V. Geometric and Photogeometric Distances to 1.47 Billion Stars in Gaia Early Data Release 3}",
      journal = {\aj},
         year = 2021,
        month = mar,
       volume = {161},
       number = {3},
          eid = {147},
        pages = {147},
          doi = {10.3847/1538-3881/abd806},
archivePrefix = {arXiv},
       eprint = {2012.05220},
 primaryClass = {astro-ph.SR},
       adsurl = {https://ui.adsabs.harvard.edu/abs/2021AJ....161..147B}
}

@BOOK{Binney2008,
       author = {{Binney}, James and {Tremaine}, Scott},
        title = "{Galactic Dynamics: Second Edition}",
    publisher = {Princeton University Press},
         year = 2008,
       adsurl = {https://ui.adsabs.harvard.edu/abs/2008gady.book.....B}
}

@ARTICLE{Bland-Hawthorn2016,
       author = {{Bland-Hawthorn}, Joss and {Gerhard}, Ortwin},
        title = "{The Galaxy in Context: Structural, Kinematic, and Integrated Properties}",
      journal = {\araa},
         year = 2016,
        month = sep,
       volume = {54},
        pages = {529-596},
          doi = {10.1146/annurev-astro-081915-023441},
archivePrefix = {arXiv},
       eprint = {1602.07702},
 primaryClass = {astro-ph.GA},
       adsurl = {https://ui.adsabs.harvard.edu/abs/2016ARA&A..54..529B}
}

@ARTICLE{Cantat-Gaudin2021,
       author = {{Cantat-Gaudin}, Tristan and {Brandt}, Timothy D.},
        title = "{Characterizing and correcting the proper motion bias of the bright Gaia EDR3 sources}",
      journal = {\aap},
         year = 2021,
        month = may,
       volume = {649},
          eid = {A124},
        pages = {A124},
          doi = {10.1051/0004-6361/202140807},
archivePrefix = {arXiv},
       eprint = {2103.07432},
 primaryClass = {astro-ph.GA},
       adsurl = {https://ui.adsabs.harvard.edu/abs/2021A&A...649A.124C}
}

@ARTICLE{Chrobakova2020,
       author = {{Chrob{\'a}kov{\'a}}, {\v{Z}}. and {L{\'o}pez-Corredoira}, M. and {Sylos Labini}, F. and {Wang}, H.-F. and {Nagy}, R.},
        title = "{Gaia-DR2 extended kinematical maps. III. Rotation curves analysis, dark matter, and MOND tests}",
      journal = {\aap},
         year = 2020,
        month = oct,
       volume = {642},
          eid = {A95},
        pages = {A95},
          doi = {10.1051/0004-6361/202038736},
archivePrefix = {arXiv},
       eprint = {2007.14825},
 primaryClass = {astro-ph.GA},
       adsurl = {https://ui.adsabs.harvard.edu/abs/2020A&A...642A..95C}
}

@ARTICLE{Denyshchenko2024,
       author = {{Denyshchenko}, S.~I. and {Fedorov}, P.~N. and {Akhmetov}, V.~S. and {Velichko}, A.~B. and {Dmytrenko}, A.~M.},
        title = "{Determining the parameters of the spiral arms of the Galaxy from kinematic tracers based on Gaia DR3 data}",
      journal = {\mnras},
         year = 2024,
        month = jan,
       volume = {527},
       number = {1},
        pages = {1472-1480},
          doi = {10.1093/mnras/stad3350},
archivePrefix = {arXiv},
       eprint = {2308.14021},
 primaryClass = {astro-ph.GA},
       adsurl = {https://ui.adsabs.harvard.edu/abs/2024MNRAS.527.1472D}
}

@ARTICLE{Ding2025,
       author = {{Ding}, Ye and {Liao}, Shilong and {Wen}, Shangyu and {Qi}, Zhaoxiang},
        title = "{Analysis of the Gaia Data Release 3 Parallax Bias at Bright Magnitudes}",
      journal = {\aj},
         year = 2025,
        month = apr,
       volume = {169},
       number = {4},
          eid = {211},
        pages = {211},
          doi = {10.3847/1538-3881/adba44},
archivePrefix = {arXiv},
       eprint = {2502.08068},
 primaryClass = {astro-ph.SR},
       adsurl = {https://ui.adsabs.harvard.edu/abs/2025AJ....169..211D}
}

@ARTICLE{Dmytrenko2025,
       author = {{Dmytrenko}, A.~M. and {Fedorov}, P.~N. and {Akhmetov}, V.~S. and {Velichko}, A.~B. and {Denyshchenko}, S.~I. and {Khramtsov}, V.~P. and {Vavilova}, I.~B. and {Dobrycheva}, D.~V. and {Sergijenko}, O. and {Kompaniiets}, A.~A. Vasylenko O.~V.},
        title = "{Spatial orientation and shape of the velocity ellipsoids of the Gaia DR3 giants and sub-giants in the Galactic plane}",
      journal = {\mnras},
         year = 2025,
        month = sep,
       volume = {542},
       number = {3},
        pages = {2542-2559},
          doi = {10.1093/mnras/staf1408},
archivePrefix = {arXiv},
       eprint = {2412.18333},
 primaryClass = {astro-ph.GA},
       adsurl = {https://ui.adsabs.harvard.edu/abs/2025MNRAS.542.2542D}
}

@ARTICLE{Dmytrenko2023,
       author = {{Dmytrenko}, A.~M. and {Fedorov}, P.~N. and {Akhmetov}, V.~S. and {Velichko}, A.~B. and {Denyshchenko}, S.~I.},
        title = "{The vertex coordinates of the Galaxy's stellar systems according to the Gaia DR3 catalogue}",
      journal = {\mnras},
         year = 2023,
        month = may,
       volume = {521},
       number = {3},
        pages = {4247-4256},
          doi = {10.1093/mnras/stad823},
archivePrefix = {arXiv},
       eprint = {2301.00203},
 primaryClass = {astro-ph.GA},
       adsurl = {https://ui.adsabs.harvard.edu/abs/2023MNRAS.521.4247D}
}

@ARTICLE{Drimmel2023,
       author = {{Gaia Collaboration} and {Drimmel}, R. and {Romero-G{\'o}mez}, M. and {Chemin}, L. and {Ramos}, P. and {Poggio}, E. and {Ripepi}, V. and {Andrae}, R. and {Blomme}, R. and {Cantat-Gaudin}, T. and {Castro-Ginard}, A. and {Clementini}, G. and {Figueras}, F. and {Fouesneau}, M. and {Fr{\'e}mat}, Y. and {Jardine}, K. and {Khanna}, S. and {Lobel}, A. and {Marshall}, D.~J. and {Muraveva}, T. and {Brown}, A.~G.~A. and {Vallenari}, A. and {Prusti}, T. and {de Bruijne}, J.~H.~J. and {Arenou}, F. and {Babusiaux}, C. and {Biermann}, M. and {Creevey}, O.~L. and {Ducourant}, C. and {Evans}, D.~W. and {Eyer}, L. and {Guerra}, R. and {Hutton}, A. and {Jordi}, C. and {Klioner}, S.~A. and {Lammers}, U.~L. and {Lindegren}, L. and {Luri}, X. and {Mignard}, F. and {Panem}, C. and {Pourbaix}, D. and {Randich}, S. and {Sartoretti}, P. and {Soubiran}, C. and {Tanga}, P. and {Walton}, N.~A. and {Bailer-Jones}, C.~A.~L. and {Bastian}, U. and {Jansen}, F. and {Katz}, D. and {Lattanzi}, M.~G. and {van Leeuwen}, F. and {Bakker}, J. and {Cacciari}, C. and {Casta{\~n}eda}, J. and {De Angeli}, F. and {Fabricius}, C. and {Galluccio}, L. and {Guerrier}, A. and {Heiter}, U. and {Masana}, E. and {Messineo}, R. and {Mowlavi}, N. and {Nicolas}, C. and {Nienartowicz}, K. and {Pailler}, F. and {Panuzzo}, P. and {Riclet}, F. and {Roux}, W. and {Seabroke}, G.~M. and {Sordo}, R. and {Th{\'e}venin}, F. and {Gracia-Abril}, G. and {Portell}, J. and {Teyssier}, D. and {Altmann}, M. and {Audard}, M. and {Bellas-Velidis}, I. and {Benson}, K. and {Berthier}, J. and {Burgess}, P.~W. and {Busonero}, D. and {Busso}, G. and {C{\'a}novas}, H. and {Carry}, B. and {Cellino}, A. and {Cheek}, N. and {Damerdji}, Y. and {Davidson}, M. and {de Teodoro}, P. and {Nu{\~n}ez Campos}, M. and {Delchambre}, L. and {Dell'Oro}, A. and {Esquej}, P. and {Fern{\'a}ndez-Hern{\'a}ndez}, J. and {Fraile}, E. and {Garabato}, D. and {Garc{\'\i}a-Lario}, P. and {Gosset}, E. and {Haigron}, R. and {Halbwachs}, J. -L. and {Hambly}, N.~C. and {Harrison}, D.~L. and {Hern{\'a}ndez}, J. and {Hestroffer}, D. and {Hodgkin}, S.~T. and {Holl}, B. and {Jan{\ss}en}, K. and {Jevardat de Fombelle}, G. and {Jordan}, S. and {Krone-Martins}, A. and {Lanzafame}, A.~C. and {L{\"o}ffler}, W. and {Marchal}, O. and {Marrese}, P.~M. and {Moitinho}, A. and {Muinonen}, K. and {Osborne}, P. and {Pancino}, E. and {Pauwels}, T. and {Recio-Blanco}, A. and {Reyl{\'e}}, C. and {Riello}, M. and {Rimoldini}, L. and {Roegiers}, T. and {Rybizki}, J. and {Sarro}, L.~M. and {Siopis}, C. and {Smith}, M. and {Sozzetti}, A. and {Utrilla}, E. and {van Leeuwen}, M. and {Abbas}, U. and {{\'A}brah{\'a}m}, P. and {Abreu Aramburu}, A. and {Aerts}, C. and {Aguado}, J.~J. and {Ajaj}, M. and {Aldea-Montero}, F. and {Altavilla}, G. and {{\'A}lvarez}, M.~A. and {Alves}, J. and {Anders}, F. and {Anderson}, R.~I. and {Anglada Varela}, E. and {Antoja}, T. and {Baines}, D. and {Baker}, S.~G. and {Balaguer-N{\'u}{\~n}ez}, L. and {Balbinot}, E. and {Balog}, Z. and {Barache}, C. and {Barbato}, D. and {Barros}, M. and {Barstow}, M.~A. and {Bartolom{\'e}}, S. and {Bassilana}, J. -L. and {Bauchet}, N. and {Becciani}, U. and {Bellazzini}, M. and {Berihuete}, A. and {Bernet}, M. and {Bertone}, S. and {Bianchi}, L. and {Binnenfeld}, A. and {Blanco-Cuaresma}, S. and {Boch}, T. and {Bombrun}, A. and {Bossini}, D. and {Bouquillon}, S. and {Bragaglia}, A. and {Bramante}, L. and {Breedt}, E. and {Bressan}, A. and {Brouillet}, N. and {Brugaletta}, E. and {Bucciarelli}, B. and {Burlacu}, A. and {Butkevich}, A.~G. and {Buzzi}, R. and {Caffau}, E. and {Cancelliere}, R. and {Carballo}, R. and {Carlucci}, T. and {Carnerero}, M.~I. and {Carrasco}, J.~M. and {Casamiquela}, L. and {Castellani}, M. and {Chaoul}, L. and {Charlot}, P. and {Chiaramida}, V. and {Chiavassa}, A. and {Chornay}, N. and {Comoretto}, G. and {Contursi}, G. and {Cooper}, W.~J. and {Cornez}, T. and {Cowell}, S. and {Crifo}, F. and {Cropper}, M. and {Crosta}, M. and {Crowley}, C. and {Dafonte}, C. and {Dapergolas}, A. and {David}, P. and {de Laverny}, P. and {De Luise}, F. and {De March}, R. and {De Ridder}, J. and {de Souza}, R. and {de Torres}, A. and {del Peloso}, E.~F. and {del Pozo}, E. and {Delbo}, M. and {Delgado}, A. and {Delisle}, J. -B. and {Demouchy}, C. and {Dharmawardena}, T.~E. and {Di Matteo}, P. and {Diakite}, S. and {Diener}, C. and {Distefano}, E. and {Dolding}, C. and {Enke}, H. and {Fabre}, C. and {Fabrizio}, M. and {Faigler}, S. and {Fedorets}, G. and {Fernique}, P. and {Fournier}, Y. and {Fouron}, C. and {Fragkoudi}, F. and {Gai}, M. and {Garcia-Gutierrez}, A. and {Garcia-Reinaldos}, M. and {Garc{\'\i}a-Torres}, M. and {Garofalo}, A. and {Gavel}, A. and {Gavras}, P. and {Gerlach}, E. and {Geyer}, R. and {Giacobbe}, P. and {Gilmore}, G. and {Girona}, S. and {Giuffrida}, G. and {Gomel}, R. and {Gomez}, A. and {Gonz{\'a}lez-N{\'u}{\~n}ez}, J. and {Gonz{\'a}lez-Santamar{\'\i}a}, I. and {Gonz{\'a}lez-Vidal}, J.~J. and {Granvik}, M. and {Guillout}, P. and {Guiraud}, J. and {Guti{\'e}rrez-S{\'a}nchez}, R. and {Guy}, L.~P. and {Hatzidimitriou}, D. and {Hauser}, M. and {Haywood}, M. and {Helmer}, A. and {Helmi}, A. and {Sarmiento}, M.~H. and {Hidalgo}, S.~L. and {H{\l}adczuk}, N. and {Hobbs}, D. and {Holland}, G. and {Huckle}, H.~E. and {Jasniewicz}, G. and {Jean-Antoine Piccolo}, A. and {Jim{\'e}nez-Arranz}, {\'O}. and {Juaristi Campillo}, J. and {Julbe}, F. and {Karbevska}, L. and {Kervella}, P. and {Kordopatis}, G. and {Korn}, A.~J. and {K{\'o}sp{\'a}l}, {\'A}. and {Kostrzewa-Rutkowska}, Z. and {Kruszy{\'n}ska}, K. and {Kun}, M. and {Laizeau}, P. and {Lambert}, S. and {Lanza}, A.~F. and {Lasne}, Y. and {Le Campion}, J. -F. and {Lebreton}, Y. and {Lebzelter}, T. and {Leccia}, S. and {Leclerc}, N. and {Lecoeur-Taibi}, I. and {Liao}, S. and {Licata}, E.~L. and {Lindstr{\o}m}, H.~E.~P. and {Lister}, T.~A. and {Livanou}, E. and {Lorca}, A. and {Loup}, C. and {Madrero Pardo}, P. and {Magdaleno Romeo}, A. and {Managau}, S. and {Mann}, R.~G. and {Manteiga}, M. and {Marchant}, J.~M. and {Marconi}, M. and {Marcos}, J. and {Marcos Santos}, M.~M.~S. and {Mar{\'\i}n Pina}, D. and {Marinoni}, S. and {Marocco}, F. and {Martin Polo}, L. and {Mart{\'\i}n-Fleitas}, J.~M. and {Marton}, G. and {Mary}, N. and {Masip}, A. and {Massari}, D. and {Mastrobuono-Battisti}, A. and {Mazeh}, T. and {McMillan}, P.~J. and {Messina}, S. and {Michalik}, D. and {Millar}, N.~R. and {Mints}, A. and {Molina}, D. and {Molinaro}, R. and {Moln{\'a}r}, L. and {Monari}, G. and {Mongui{\'o}}, M. and {Montegriffo}, P. and {Montero}, A. and {Mor}, R. and {Mora}, A. and {Morbidelli}, R. and {Morel}, T. and {Morris}, D. and {Murphy}, C.~P. and {Musella}, I. and {Nagy}, Z. and {Noval}, L. and {Oca{\~n}a}, F. and {Ogden}, A. and {Ordenovic}, C. and {Osinde}, J.~O. and {Pagani}, C. and {Pagano}, I. and {Palaversa}, L. and {Palicio}, P.~A. and {Pallas-Quintela}, L. and {Panahi}, A. and {Payne-Wardenaar}, S. and {Pe{\~n}alosa Esteller}, X. and {Penttil{\"a}}, A. and {Pichon}, B. and {Piersimoni}, A.~M. and {Pineau}, F. -X. and {Plachy}, E. and {Plum}, G. and {Pr{\v{s}}a}, A. and {Pulone}, L. and {Racero}, E. and {Ragaini}, S. and {Rainer}, M. and {Raiteri}, C.~M. and {Ramos-Lerate}, M. and {Re Fiorentin}, P. and {Regibo}, S. and {Richards}, P.~J. and {Rios Diaz}, C. and {Riva}, A. and {Rix}, H. -W. and {Rixon}, G. and {Robichon}, N. and {Robin}, A.~C. and {Robin}, C. and {Roelens}, M. and {Rogues}, H.~R.~O. and {Rohrbasser}, L. and {Rowell}, N. and {Royer}, F. and {Ruz Mieres}, D. and {Rybicki}, K.~A. and {Sadowski}, G. and {S{\'a}ez N{\'u}{\~n}ez}, A. and {Sagrist{\`a} Sell{\'e}s}, A. and {Sahlmann}, J. and {Salguero}, E. and {Samaras}, N. and {Sanchez Gimenez}, V. and {Sanna}, N. and {Santove{\~n}a}, R. and {Sarasso}, M. and {Schultheis}, M.~S. and {Sciacca}, E. and {Segol}, M. and {Segovia}, J.~C. and {S{\'e}gransan}, D. and {Semeux}, D. and {Shahaf}, S. and {Siddiqui}, H.~I. and {Siebert}, A. and {Siltala}, L. and {Silvelo}, A. and {Slezak}, E. and {Slezak}, I. and {Smart}, R.~L. and {Snaith}, O.~N. and {Solano}, E. and {Solitro}, F. and {Souami}, D. and {Souchay}, J. and {Spagna}, A. and {Spina}, L. and {Spoto}, F. and {Steele}, I.~A. and {Steidelm{\"u}ller}, H. and {Stephenson}, C.~A. and {S{\"u}veges}, M. and {Surdej}, J. and {Szabados}, L. and {Szegedi-Elek}, E. and {Taris}, F. and {Taylor}, M.~B. and {Teixeira}, R. and {Tolomei}, L. and {Tonello}, N. and {Torra}, F. and {Torra}, J. and {Torralba Elipe}, G. and {Trabucchi}, M. and {Tsounis}, A.~T. and {Turon}, C. and {Ulla}, A. and {Unger}, N. and {Vaillant}, M.~V. and {van Dillen}, E. and {van Reeven}, W. and {Vanel}, O. and {Vecchiato}, A. and {Viala}, Y. and {Vicente}, D. and {Voutsinas}, S. and {Weiler}, M. and {Wevers}, T. and {Wyrzykowski}, {\L}. and {Yoldas}, A. and {Yvard}, P. and {Zhao}, H. and {Zorec}, J. and {Zucker}, S. and {Zwitter}, T.},
        title = "{Gaia Data Release 3. Mapping the asymmetric disc of the Milky Way}",
      journal = {\aap},
         year = 2023,
        month = jun,
       volume = {674},
          eid = {A37},
        pages = {A37},
          doi = {10.1051/0004-6361/202243797},
archivePrefix = {arXiv},
       eprint = {2206.06207},
 primaryClass = {astro-ph.GA},
       adsurl = {https://ui.adsabs.harvard.edu/abs/2023A&A...674A..37G}
}

@ARTICLE{Eilers2019,
       author = {{Eilers}, Anna-Christina and {Hogg}, David W. and {Rix}, Hans-Walter and {Ness}, Melissa K.},
        title = "{The Circular Velocity Curve of the Milky Way from 5 to 25 kpc}",
      journal = {\apj},
         year = 2019,
        month = jan,
       volume = {871},
       number = {1},
          eid = {120},
        pages = {120},
          doi = {10.3847/1538-4357/aaf648},
archivePrefix = {arXiv},
       eprint = {1810.09466},
 primaryClass = {astro-ph.GA},
       adsurl = {https://ui.adsabs.harvard.edu/abs/2019ApJ...871..120E}
}

@ARTICLE{Fedorov2023,
       author = {{Fedorov}, P.~N. and {Akhmetov}, V.~S. and {Velichko}, A.~B. and {Dmytrenko}, A.~M. and {Denyshchenko}, S.~I.},
        title = "{Mapping the kinematic parameters of the Galaxy from the Gaia EDR3 red giants and sub-giants}",
      journal = {\mnras},
         year = 2023,
        month = jan,
       volume = {518},
       number = {2},
        pages = {2761-2774},
          doi = {10.1093/mnras/stac3218},
archivePrefix = {arXiv},
       eprint = {2205.09381},
 primaryClass = {astro-ph.GA},
       adsurl = {https://ui.adsabs.harvard.edu/abs/2023MNRAS.518.2761F}
}

@ARTICLE{Jiao2023,
       author = {{Jiao}, Yongjun and {Hammer}, Fran{\c{c}}ois and {Wang}, Haifeng and {Wang}, Jianling and {Amram}, Philippe and {Chemin}, Laurent and {Yang}, Yanbin},
        title = "{Detection of the Keplerian decline in the Milky Way rotation curve}",
      journal = {\aap},
         year = 2023,
        month = oct,
       volume = {678},
          eid = {A208},
        pages = {A208},
          doi = {10.1051/0004-6361/202347513},
archivePrefix = {arXiv},
       eprint = {2309.00048},
 primaryClass = {astro-ph.GA},
       adsurl = {https://ui.adsabs.harvard.edu/abs/2023A&A...678A.208J}
}

@ARTICLE{Kompaniiets2025,
       author = {{Kompaniiets}, O.~V. and {Vavilova}, I.~B. and {Kukhar}, O.~M. and {Dobrycheva}, D.~V. and {Fedorov}, P.~N. and {Dmytrenko}, A.~M. and {Khramtsov}, V.~P. and {Sergijenko}, O.~M. and {Vasylenko}, A.~A.},
        title = "{Milky Way galaxy-analogs and isolated galaxies with bars: environmental density in the Local Volume}",
      journal = {arXiv e-prints},
         year = 2025,
        month = nov,
          eid = {arXiv:2511.18612},
        pages = {arXiv:2511.18612},
archivePrefix = {arXiv},
       eprint = {2511.18612},
 primaryClass = {astro-ph.GA},
       adsurl = {https://ui.adsabs.harvard.edu/abs/2025arXiv251118612K}
}

@ARTICLE{Koop2024,
       author = {{Koop}, Orlin and {Antoja}, Teresa and {Helmi}, Amina and {Callingham}, Thomas M. and {Laporte}, Chervin F.~P.},
        title = "{Assessing the robustness of the Galactic rotation curve inferred from the Jeans equations using Gaia DR3 and cosmological simulations}",
      journal = {\aap},
         year = 2024,
        month = dec,
       volume = {692},
          eid = {A50},
        pages = {A50},
          doi = {10.1051/0004-6361/202450911},
archivePrefix = {arXiv},
       eprint = {2405.19028},
 primaryClass = {astro-ph.GA},
       adsurl = {https://ui.adsabs.harvard.edu/abs/2024A&A...692A..50K}
}

@ARTICLE{Lindegren2021,
       author = {{Lindegren}, L. and {Bastian}, U. and {Biermann}, M. and {Bombrun}, A. and {de Torres}, A. and {Gerlach}, E. and {Geyer}, R. and {Hern{\'a}ndez}, J. and {Hilger}, T. and {Hobbs}, D. and {Klioner}, S.~A. and {Lammers}, U. and {McMillan}, P.~J. and {Ramos-Lerate}, M. and {Steidelm{\"u}ller}, H. and {Stephenson}, C.~A. and {van Leeuwen}, F.},
        title = "{Gaia Early Data Release 3. Parallax bias versus magnitude, colour, and position}",
      journal = {\aap},
         year = 2021,
        month = may,
       volume = {649},
          eid = {A4},
        pages = {A4},
          doi = {10.1051/0004-6361/202039653},
archivePrefix = {arXiv},
       eprint = {2012.01742},
 primaryClass = {astro-ph.IM},
       adsurl = {https://ui.adsabs.harvard.edu/abs/2021A&A...649A...4L}
}

@ARTICLE{Lopez-Corredoira2019,
       author = {{L{\'o}pez-Corredoira}, M. and {Sylos Labini}, F.},
        title = "{Gaia-DR2 extended kinematical maps . I. Method and application}",
      journal = {\aap},
         year = 2019,
        month = jan,
       volume = {621},
          eid = {A48},
        pages = {A48},
          doi = {10.1051/0004-6361/201833849},
archivePrefix = {arXiv},
       eprint = {1810.13436},
 primaryClass = {astro-ph.GA},
       adsurl = {https://ui.adsabs.harvard.edu/abs/2019A&A...621A..48L}
}

@ARTICLE{Lucy1974,
       author = {{Lucy}, L.~B.},
        title = "{An iterative technique for the rectification of observed distributions}",
      journal = {\aj},
         year = 1974,
        month = jun,
       volume = {79},
        pages = {745},
          doi = {10.1086/111605},
       adsurl = {https://ui.adsabs.harvard.edu/abs/1974AJ.....79..745L}
}

@ARTICLE{MaizApellaniz2021,
       author = {{Ma{\'\i}z Apell{\'a}niz}, J. and {Pantaleoni Gonz{\'a}lez}, M. and {Barb{\'a}}, R.~H.},
        title = "{Validation of the accuracy and precision of Gaia EDR3 parallaxes with globular clusters}",
      journal = {\aap},
         year = 2021,
        month = may,
       volume = {649},
          eid = {A13},
        pages = {A13},
          doi = {10.1051/0004-6361/202140418},
archivePrefix = {arXiv},
       eprint = {2101.10206},
 primaryClass = {astro-ph.IM},
       adsurl = {https://ui.adsabs.harvard.edu/abs/2021A&A...649A..13M}
}

@ARTICLE{MaizApellaniz2022,
       author = {{Ma{\'\i}z Apell{\'a}niz}, J.},
        title = "{An estimation of the Gaia EDR3 parallax bias from stellar clusters and Magellanic Clouds data}",
      journal = {\aap},
         year = 2022,
        month = jan,
       volume = {657},
          eid = {A130},
        pages = {A130},
          doi = {10.1051/0004-6361/202142365},
archivePrefix = {arXiv},
       eprint = {2110.01475},
 primaryClass = {astro-ph.IM},
       adsurl = {https://ui.adsabs.harvard.edu/abs/2022A&A...657A.130M}
}

@ARTICLE{McMillan2022,
       author = {{McMillan}, Paul J. and {Petersson}, Jonathan and {Tepper-Garcia}, Thor and {Bland-Hawthorn}, Joss and {Antoja}, Teresa and {Chemin}, Laurent and {Figueras}, Francesca and {Khanna}, Shourya and {Kordopatis}, Georges and {Ramos}, Pau and {Romero-G{\'o}mez}, Merce and {Seabroke}, George},
        title = "{The disturbed outer Milky Way disc}",
      journal = {\mnras},
         year = 2022,
        month = nov,
       volume = {516},
       number = {4},
        pages = {4988-5002},
          doi = {10.1093/mnras/stac2571},
archivePrefix = {arXiv},
       eprint = {2206.04059},
 primaryClass = {astro-ph.GA},
       adsurl = {https://ui.adsabs.harvard.edu/abs/2022MNRAS.516.4988M}
}

@ARTICLE{Mroz2019,
       author = {{Mr{\'o}z}, Przemek and {Udalski}, Andrzej and {Skowron}, Dorota M. and {Skowron}, Jan and {Soszy{\'n}ski}, Igor and {Pietrukowicz}, Pawe{\l} and {Szyma{\'n}ski}, Micha{\l} K. and {Poleski}, Rados{\l}aw and {Koz{\l}owski}, Szymon and {Ulaczyk}, Krzysztof},
        title = "{Rotation Curve of the Milky Way from Classical Cepheids}",
      journal = {\apjl},
         year = 2019,
        month = jan,
       volume = {870},
       number = {1},
          eid = {L10},
        pages = {L10},
          doi = {10.3847/2041-8213/aaf73f},
archivePrefix = {arXiv},
       eprint = {1810.02131},
 primaryClass = {astro-ph.GA},
       adsurl = {https://ui.adsabs.harvard.edu/abs/2019ApJ...870L..10M}
}

@BOOK{Ogorodnikov1965,
       author = {{Ogorodnikov}, K.~F.},
        title = "{Dynamics of Stellar Systems}",
     publisher = {Pergamon Press},
         year = 1965,
       adsurl = {https://ui.adsabs.harvard.edu/abs/1965dss..book.....O}
}

@ARTICLE{Ou2024,
       author = {{Ou}, Xiaowei and {Eilers}, Anna-Christina and {Necib}, Lina and {Frebel}, Anna},
        title = "{The dark matter profile of the Milky Way inferred from its circular velocity curve}",
      journal = {\mnras},
         year = 2024,
        month = feb,
       volume = {528},
       number = {1},
        pages = {693-710},
          doi = {10.1093/mnras/stae034},
archivePrefix = {arXiv},
       eprint = {2303.12838},
 primaryClass = {astro-ph.GA},
       adsurl = {https://ui.adsabs.harvard.edu/abs/2024MNRAS.528..693O}
}

@ARTICLE{Puglisi2023,
       author = {{Puglisi}, Annagrazia and {Dudzevi{\v{c}}i{\={u}}t{\.{e}}}, Ugn{\.{e}} and {Swinbank}, Mark and {Gillman}, Steven and {Tiley}, Alfred L. and {Bower}, Richard G. and {Cirasuolo}, Michele and {Cortese}, Luca and {Glazebrook}, Karl and {Harrison}, Chris and {Ibar}, Edo and {Molina}, Juan and {Obreschkow}, Danail and {Oman}, Kyle A. and {Schaller}, Matthieu and {Shankar}, Francesco and {Sharples}, Ray M.},
        title = "{KURVS: the outer rotation curve shapes and dark matter fractions of z   1.5 star-forming galaxies}",
      journal = {\mnras},
         year = 2023,
        month = sep,
       volume = {524},
       number = {2},
        pages = {2814-2835},
          doi = {10.1093/mnras/stad1966},
archivePrefix = {arXiv},
       eprint = {2305.04382},
 primaryClass = {astro-ph.GA},
       adsurl = {https://ui.adsabs.harvard.edu/abs/2023MNRAS.524.2814P}
}

@ARTICLE{Reid2004,
       author = {{Reid}, M.~J. and {Brunthaler}, A.},
        title = "{The Proper Motion of Sagittarius A*. II. The Mass of Sagittarius A*}",
      journal = {\apj},
         year = 2004,
        month = dec,
       volume = {616},
       number = {2},
        pages = {872-884},
          doi = {10.1086/424960},
archivePrefix = {arXiv},
       eprint = {astro-ph/0408107},
 primaryClass = {astro-ph},
       adsurl = {https://ui.adsabs.harvard.edu/abs/2004ApJ...616..872R}
}

@ARTICLE{Skowron2019,
       author = {{Skowron}, Dorota M. and {Skowron}, Jan and {Mr{\'o}z}, Przemek and {Udalski}, Andrzej and {Pietrukowicz}, Pawe{\l} and {Soszy{\'n}ski}, Igor and {Szyma{\'n}ski}, Micha{\l} K. and {Poleski}, Rados{\l}aw and {Koz{\l}owski}, Szymon and {Ulaczyk}, Krzysztof and {Rybicki}, Krzysztof and {Iwanek}, Patryk},
        title = "{A three-dimensional map of the Milky Way using classical Cepheid variable stars}",
      journal = {Science},
         year = 2019,
        month = aug,
       volume = {365},
       number = {6452},
        pages = {478-482},
          doi = {10.1126/science.aau3181},
archivePrefix = {arXiv},
       eprint = {1806.10653},
 primaryClass = {astro-ph.GA},
       adsurl = {https://ui.adsabs.harvard.edu/abs/2019Sci...365..478S}
}

@ARTICLE{Vasiliev2021,
       author = {{Vasiliev}, Eugene and {Belokurov}, Vasily and {Erkal}, Denis},
        title = "{Tango for three: Sagittarius, LMC, and the Milky Way}",
      journal = {\mnras},
         year = 2021,
        month = feb,
       volume = {501},
       number = {2},
        pages = {2279-2304},
          doi = {10.1093/mnras/staa3673},
archivePrefix = {arXiv},
       eprint = {2009.10726},
 primaryClass = {astro-ph.GA},
       adsurl = {https://ui.adsabs.harvard.edu/abs/2021MNRAS.501.2279V}
}

@ARTICLE{Vavilova2024,
       author = {{Vavilova}, I.~B. and {Fedorov}, P.~M. and {Dobrycheva}, D.~V. and {Sergijenko}, O. and {Vasylenko}, A.~A. and {Dmytrenko}, A.~M. and {Khramtsov}, V.~P. and {Kompaniiets}, O.~V.},
        title = "{An advanced approach to the definition of the ``Milky Way galaxies-analogues''}",
      journal = {Space Science and Technology},
         year = 2024,
        month = jan,
       volume = {30},
       number = {4},
        pages = {81},
          doi = {10.15407/knit2024.04.081},
archivePrefix = {arXiv},
       eprint = {2410.04411},
 primaryClass = {astro-ph.GA},
       adsurl = {https://ui.adsabs.harvard.edu/abs/2024KosNT..30d..81V}
}

@ARTICLE{Wang2023,
       author = {{Wang}, Hai-Feng and {Chrob{\'a}kov{\'a}}, {\v{Z}}ofia and {L{\'o}pez-Corredoira}, Mart{\'\i}n and {Sylos Labini}, Francesco},
        title = "{Mapping the Milky Way Disk with Gaia DR3: 3D Extended Kinematic Maps and Rotation Curve to {\ensuremath{\approx}}30 kpc}",
      journal = {\apj},
         year = 2023,
        month = jan,
       volume = {942},
       number = {1},
          eid = {12},
        pages = {12},
          doi = {10.3847/1538-4357/aca27c},
archivePrefix = {arXiv},
       eprint = {2211.05668},
 primaryClass = {astro-ph.GA},
       adsurl = {https://ui.adsabs.harvard.edu/abs/2023ApJ...942...12W}
}

@ARTICLE{Zhou2023,
       author = {{Zhou}, Yuan and {Li}, Xinyi and {Huang}, Yang and {Zhang}, Huawei},
        title = "{The Circular Velocity Curve of the Milky Way from 5-25 kpc Using Luminous Red Giant Branch Stars}",
      journal = {\apj},
         year = 2023,
        month = apr,
       volume = {946},
       number = {2},
          eid = {73},
        pages = {73},
          doi = {10.3847/1538-4357/acadd9},
archivePrefix = {arXiv},
       eprint = {2212.10393},
 primaryClass = {astro-ph.GA},
       adsurl = {https://ui.adsabs.harvard.edu/abs/2023ApJ...946...73Z}
}

\begin{appendix}

\end{appendix}


\end{document}